\documentclass{jpp}
\usepackage{graphicx}
\usepackage[dvipsname]{xcolor}
\usepackage{subfigure}
\usepackage{bm}
\usepackage[outdir=./]{epstopdf}
\usepackage[colorlinks=true, linkcolor=olive,citecolor=olive]{hyperref}

\shortauthor{J.-M. Rax and R. Gueroult}

\title{Faraday-Fresnel rotation and splitting of orbital angular momentum carrying waves in a rotating plasma}

\author{J.-M. Rax\aff{1}
 \and R. Gueroult\aff{2}}

\affiliation{\aff{1}IJCLab - Universit\'{e} de Paris-Saclay \& D{\'e}partement de Physique - Ecole Polytechnique, 91405 Orsay, France
\aff{2}LAPLACE, Universit\'{e} de Toulouse, CNRS, INPT, UPS, 31062 Toulouse, France}

\begin{document}

\maketitle

\begin{abstract}
Rotational Fresnel drag - or orbital Faraday rotation - in a rotating magnetised plasma is uncovered and studied analytically for Trivelpiece-Gould and Whistler-Helicon waves carrying orbital angular momentum (OAM). Plasma rotation is shown to introduce a non-zero phase shift between OAM-carrying eigenmodes with opposite helicities, similarly to the phase-shift between spin angular momentum eigenmodes associated with the classical Faraday effect in a magnetised plasma at rest. By examining the dispersion relation for these two low-frequency modes in a Brillouin rotating plasma, this Faraday-Fresnel rotation effect is traced back to the combined effects of Doppler shift, centrifugal forces and Coriolis forces. In addition, rotation is further shown to lead to rotation- and azimuthal mode-dependent longitudinal group velocity, therefore prediciting the Faraday-Fresnel splitting of the enveloppe of a wave packet containing a superposition of OAM-carrying eigenmodes with opposite helicities.


\end{abstract}

\section{Introduction}

The dynamics of magnetised plasmas and beams confined by an
axial magnetic field and put in rotation through a radial electric field has been
extensively investigated, both from a basic plasma physics standpoint and to enable applications. Rotating plasmas have for instance been investigated - and continue to - for the purpose of magnetic confinement fusion~\citep{Wilcox1959}, including in homopolar configurations~\citep{Anderson1958}, rotating mirrors~\citep{Lehnert1971,Bekhtenev1980,Ellis2005,Fetterman2008,Fetterman2010} and rotating tokamaks~\citep{Stix1971,Rax2017,Ochs2017}. Rotating plasmas have also been the object of continued attention for mass separation applications~\citep{Dolgolenko2009,Dolgolenko2017,Zweben2018}, notably in the form of plasma centrifuges~\citep{Bonnevier1966,Krishnan1981,Prasad1987}, crossed-field rotating plasmas~\citep{Ohkawa2002,Shinohara2007,Gueroult2019,Liziakin2021} and rotating magnetic field configurations~\citep{Rax2016}. In parallel to these applications driven studies, the complementary problem of angular momentum coupling between electromagnetic waves and particles has been examined in the study of particle acceleration~\citep{Rax2010,Thaury2013}, magnetic field generation in plasma channels and plasma bubbles~\citep{Shvets2002,Kostyukov2002} and isotope separation~\citep{Rax2007}.

A basic question of importance to these studies, but also to shed light on the intricate effects of rotation  in astrophysics~\citep{Balbus1998,Benomar2018}, is how to diagnose plasma rotation. Asking this question is fundamentally asking for physical probes sensitive to fluid vorticity. Interestingly, it has recently been shown that rotation could imprint its signature on the polarisation of a wave propagating through a rotating plasma~\citep{Gueroult2019a}, offering in turn unique opportunities to infer the rotation direction in pulsars.  A challenge here, however, is that in a magnetised plasma both this polarisation drag effect~\citep{Player1976,Jones1976} (also referred to as mechanical Faraday effect) and the classic Faraday rotation~\citep{Faraday1846,Stix1992} are at play~\citep{Gueroult2020}. Measuring plasma rotation through polarisation drag hence requires disambiguating this effect from Faraday rotation. Similarly, harnessing polarisation drag in laboratory experiments likely requires conditions such that polarisation drag dominates over Faraday rotation~\citep{Gueroult2020}. Polarisation however corresponds only to the spin component of the wave angular momentum. Indeed, waves can also carry orbital angular momentum through helical phase fronts of the type $\exp j\left( l\theta -\beta z\right)$ with $l\in \Bbb{Z}$ and $\beta \in \Bbb{R}$, and where $\theta$ and $z$ are the transverse azimuthal angle and the propagation axis in a cylindrical coordinate system~\citep{Gough1986,Allen1992,Enk1994,Barnett1994,Stenzel2016a,Barnett2017}.

In isotropic media mechanical rotation has been postulated~\citep{Padgett2006,Goette2007,Leach2008} and then demonstrated experimentally~\citep{Franke-Arnold2011} to lead to image rotation. This phenomena, analogously to the polarisation rotation that stems from a phase shift between right- and left-circularly polarised modes, results from a mechanically induced phase shift between the positive and negative OAM carrying modes~\citep{Wisniewski-Barker2014}. Image rotation has hence been coined as mechanical Faraday effect for OAM-carrying beams. Yet, owing to its similarities with Fresnel drag for a uniform motion~\citep{Fresnel1818}, image rotation is equally referred to as rotatory photon drag~\citep{Franke-Arnold2011}. Reflecting these two roots, we refer to it here as Faraday-Fresnel rotation (FFR).

An interesting property of FFR in a dielectric medium is that, in contrast to polarisation rotation, it is insensitive to a magnetic field aligned with the propagation direction.  Measuring FFR could hence lift the challenges faced by polarisation drag to measure rotation in a rotating magnetised plasma, as suggested for ultracold atomic gases~\citep{Ruseckas2007}. Yet, while both linear and angular Fresnel drag have been studied in isotropic, non-gyrotropic rotating media~\citep{Goette2007,Leach2008}, the case of rotating gyrotropic media such as a rotating magnetised plasma has not yet been investigated. Conceptually, anisotropy and gyrotropy bring out additional complexity in that the polarisation state and the wave vector direction can be set independently for a given frequency in isotropic media, whereas these two parameters are no longer independent in anisotropic or gyrotropic media. 

Two approches can be used to examine the effect of rotation on waves carrying OAM in a rotating magnetised plasma. One possibility is to consider the transformation of the various parameters from the rotating plasma rest frame to the lab frame. This is the approach we previously used to uncover the effect of plasma rotation on the wave's spin angular momentum~\citep{Gueroult2019a,Gueroult2020}. Another possibility is to carry out the calculations in the lab frame starting from first principles. This is the approach we will use in this work as it proves more straightforward and also provides more directly physical insights into this new plasma effect. The approach in this study is therefore to solve the particles dynamics to calculate the rotating plasma linear response. We specifically consider here the low frequency electronic branches where the coupling between the fields and the particles is strong, that is Trivelpiece-Gould (TG) and Whistler-Helicon (WH) modes~\citep{Stix1992,Davidson2001}.

This paper is organised as follows. In the next section, Section~\ref{Sec:II}, we introduce the effect of rotation on plasma waves carrying OAM through an analogy with SAM-carrying waves in the presence of a magnetic field, and briefly recall the characteritics of Trivelpiece-Gould (TG) and Whistler-Helicon (WH) plasma waves as they will be the focus of our work. Section~\ref{Sec:III} then summarises the properties of Brillouin rotation in a cold magnetised plasma column used later on as our rotating plasma equilibrium. With these tools in hand, the dispersion relation for TG and WH waves in a rotating plasma are derived and analysed in Sections~\ref{Sec:IV} and \ref{Sec:V}, respectively. In both instances Faraday-Fresnel rotation and its parametric dependencies are identified and antennae designs that could excite these rotating waves are proposed. Building on this finding, Section~\ref{Sec:VI} examines the effect on a wave packet containing a combination of counter rotating OAM-carrying eigenmodes, and shows how rotation lead to the Faraday-Fresnel splitting of this wave packet. Finally Section~\ref{Sec:Conclusion} summarises the main findings of this study.

\section{Phenomenology of Faraday-Fresnel rotation in a rotating plasma}
\label{Sec:II}

Throughout this study we write $\left( r,\theta ,z\right)$ a set of cylindrical coordinates on the cylindrical coordinate system $\left( \mathbf{e}_{r},\mathbf{e}_{\theta },\mathbf{e}_{z}\right)$, and define $z$ as the direction of both the background static magnetic field $\mathbf{B}_0 = B_0\mathbf{e}_{z}$ and the plasma angular velocity $\bm{\Omega}=\Omega\mathbf{e}_{z}$.

\subsection{Polarisation rotation from phase shift between circularly polarised eigenmodes}

To draw an analogy between the classical Faraday effect and Faraday-Fresnel rotation, let us first recall the basic picture relating polarisation rotation to circularly polarised eigenmodes. For this consider a plane wave with a phase factor $\exp j\left( \omega t-\beta z\right) $ propagating along the magnetic field in an homogeneous plasma at rest. The study of the cold plasma dispersion relation between the axial wave vector $\beta $ and the frequency $\omega $ leads to the identification of the two left (L) and right (R) circularly polarised modes such that 
\begin{equation}
\beta _{L/R}=\frac{\omega }{c}\sqrt{\frac{\left( \omega \pm \omega
_{L}\right) \left( \omega \mp \omega _{R}\right) }{\left( \omega \mp \omega
_{ce}\right) \left( \omega \pm \omega _{ci}\right) }}.  \label{d1}
\end{equation}
Here $\omega_{c\alpha} = |q_{\alpha}|B_0/m_{\alpha}$, $\alpha=i,e$ is the (positive) cyclotron frequency and $\omega_{L/R} = ({\omega_{pe}}^2+{\omega_{ce}}^2/4)^{1/2}\pm\omega_{ce}/2$ are the left and right cut-off with $\omega_{pe} = ne^2/(m_e\epsilon_0)$ the plasma frequency. At high frequency $\omega >\omega _{ce}$ the electrons inertia response dominates the dispersion relation and Eq. (\ref{d1}) can be expanded to give the phase difference $\delta \varphi $ between the $L$ and $R$ modes accumulated when propagating one wavelength $\lambda = (\lambda_R+\lambda_L)/2$ along $z$ 
\begin{equation}
\left. \frac{\delta \varphi }{4\pi }\right| _{SAM}=\frac{\beta _{L}-\beta
_{R}}{\beta _{L}+\beta _{R}}\approx \frac{\omega _{ce}}{\omega }\frac{\omega
_{pe}^{2}}{\omega ^{2}}.  \label{d2}
\end{equation}
The phase difference derived in Eq. (\ref{d2}) is the classical formula describing the high-frequency Faraday SAM rotation in a magnetised plasma at rest. A very similar result, though not with the same frequency scaling, can be obtained by considering the phase velocity difference between right- and left-circularly polarised wave as a result of the medium's rotation~\citep{Player1976,Jones1976,Goette2007,Gueroult2020}.

\subsection{Faraday-Fresnel rotation from phase shift between OAM-carrying eigenmodes}

Plasma waves carrying orbital angular momentum (see Appendix~\ref{Sec:Appendix_A}  for a discussion of spin and orbital angular momentum in waves and particles) have been the object of growing attention over the last decade. In unmagnetised plasmas, OAM-carrying plasma waves have been studied theoretically mostly under the scalar paraxial approximation (see, \emph{e.~g.}, \cite{Mendonca2009,Mendonca2012,Mendonca2012a}), although exact solutions of the vector Maxwell equations have recently been derived~\citep{Chen2017}. Meanwhile, in cold magnetised plasmas, waves carrying orbital angular momentum have been reported in the form of Trivelpiece-Gould (TG) and Whistler-Helicon (WH) waves~\citep{Stenzel2015a,Urrutia2016,Stenzel2016a}, as well as predicted for twisted shear Alfv{\'e}n waves~\citep{Shukla2012}. Because our interest in this study, as it will be detailed in the next section, is in Brillouin rotating plasmas, we focus our attention on magnetised plasmas and consider more specifically TG and WH modes. 

Whistler-Helicon (WH) waves are intermediate frequencies $\omega _{lh}<\omega <\omega _{ce}$ waves propagating along the magnetic field, with $\omega_{lh}$ the lower hybrid frequency. The plasma response in this regime is dominated by the electron Hall current, which allows simplifying the dispersion relation Eq. (\ref{d1}) to obtain the classical Whistler dispersion 
\begin{equation}
{\beta _{W}}^{2}\approx \frac{\omega }{c^{2}}\frac{\omega _{L}\omega _{R}}{%
\omega _{ce}}\approx \frac{{\omega_{pe}}^{2}}{c^{2}}\frac{\omega }{\omega
_{ce}}.  \label{dw}
\end{equation}
On the other hand non-rotating Trivelpiece-Gould (TG) are electrostatic modes that propagate near the electron cyclotron frequency $\omega _{ce}$ and the electron plasma frequency $\omega _{pe}$. These waves are characterised by an electric potential proportional to $J_{l}\left( \beta _{\bot }r\right) \exp j\left( \omega t-\beta _{TG}z\right) $ where $J_{l}$ is the ordinary Bessel function of order $l\in \Bbb{Z}$, and the dispersion relation in this case is~\citep{Trivelpiece1959}
\begin{equation}
{\beta _{TG}}^{2}=\beta _{\bot }^{2}\frac{\omega ^{2}\left({\omega_{uh}}^{2}-\omega ^{2}\right) }{\left( {\omega_{ce}}^{2}-\omega ^{2}\right)
\left( {\omega_{pe}}^{2}-\omega ^{2}\right) }  \label{dt}
\end{equation}
with $\omega_{uh}\sim({\omega_{pe}}^2+{\omega_{ce}}^2)^{1/2}$ the upper hybrid frequency.



As mentioned above, both TG waves~\citep{Urrutia2016} and WH waves~\citep{Stenzel2015a} with helical phase front of the form $\exp j\left(\omega t\pm l\theta -\beta_{\pm l}z\right)$ can propagate in a cold magnetised plasma at rest~\citep{Stenzel2016a}. Both positive and negative azimuthal mode numbers $l$, that is to say both positive and negative orbital helicities for the same absolute value of OAM content, can propagate and be generated through an appropriate phasing of the generating antenna array. It is important to underline here though that the wave rotation should not be construed as the rotation of the field vectors (\emph{i.~e.} the waves polarisation), and only indicates a rotation of the transverse wave pattern. Whistler modes are for instance right circularly polarised modes.  The superposition of azimuthal modes of equal amplitudes but opposite azimuthal mode number (OAM of $\pm l\hbar$) then leads to an azimuthally standing wave (zero OAM), very much like the superposition of right- and left-circularly polarised modes (SAM of $\pm\hbar$) leads to a linearly polarised wave (zero SAM). 

By analogy with the phase shift between opposite SAM eigenmodes given in Eq.~(\ref{d2}), we can define the phase shift between opposite OAM-carrying eigenmodes
\begin{equation}
\left. \frac{\delta \varphi }{4\pi }\right| _{OAM}=\left. \frac{\beta
_{+l}-\beta _{-l}}{\beta _{+l}+\beta _{-l}}\right| _{TG/W}\approx l\frac{%
\Omega }{\omega }F_{TG/W}\left( \omega _{pe},\omega _{ce},\omega \right). \label{dtg}
\end{equation}
As we will show, this OAM phase difference is zero absent rotation (\emph{i.e.} for $\Omega=0$) for both WH and TG modes. This is because in this case $\beta_{+l} = \beta_{-}$, so that the transverse wave pattern produced by the superposition of opposite OAM-carrying eigenmodes does not depend on $z$. This is analog to the constant linear polarisation observed if the phase velocity of right- and left-circularly polarised modes is the same. Building on this analogy, we will show that plasma rotation $\Omega\neq0$ leads to $\beta_{+l} \neq \beta_{-}$, and therefore now to a rotation of the transverse wave pattern along $z$, which we refer to as Faraday-Fresnel rotation. The dispersive form factor $F$ introduced in Eq.~(\ref{dtg}) and derived in Eq.~(\ref{ff1}) and Eq.~(\ref{ff2}) for TG and WH waves, respectively, quantifies the strength of this effect.

Finally, note that this study is limited to the analysis of the OAM dynamics through the reactive (or adiabatic) part of the plasma response. The active (or resonant) part of the plasma response, which leads to SAM and OAM transfer, is assumed negligible as we consider a cold collisionless plasma model. This active or resonant part of the interaction is only briefly discussed in Appendix~\ref{Sec:Appendix_A} to provide a fuller picture of the SAM and OAM dynamics.

\section{Brillouin rotation of a magnetised plasma column}
\label{Sec:III}

Before embarking on the derivation of Faraday-Fresnel rotation for the Trivelpiece-Gould and Whistler-Helicon modes, the properties of the Brillouin rotating plasmas considered in the rest of this manuscript are briefly recalled here. 

Brillouin rotation refers to the rotation of a magnetised plasma column which results from the combination of an axial confining magnetic field $\mathbf{B}_{0}=B_{0}\mathbf{e}_{z}$ and a radial electric field $\mathbf{E}_{0}=B_{0}\Omega r\mathbf{e}_{r}$. As shown next the plasma angular velocity $\Omega =d\theta /dt$ is then the $E\times B$ drift rigid body rotation corrected by inertial effects~\citep{Brillouin1945}, as well as possibly by collisional effects~\citep{Rax2015,Rax2019a}. In a non-neutral plasma the electric field $\mathbf{E}_{0}$ is the space charge field~\citep{Davidson2001}. On the other hand, the electric field $\mathbf{E}_{0}$ in a quasi-neutral plasma requires in steady-state maintaining a small uniform space charge $\varepsilon _{0}\bm{\nabla }\cdot \mathbf{E}_{0}$ = $%
2\varepsilon _{0}\Omega B_{0}$, and a DC or RF power input is thus needed to maintain this small deviation from quasi-neutrality against short circuiting radial currents~\citep{Rax2019a,Kolmes2019}.

Neglecting for simplicity collisions in this study and writing $\mathbf{v}_{0}$ the particle fluid velocity, the steady state momentum balance between inertia, electric and magnetic forces writes
\begin{equation}
m\left( \mathbf{v}_{0}\cdot \bm{\nabla }\right) \mathbf{v}_{0}=q\left( 
\mathbf{E}_{0}+\mathbf{v}_{0}\times \mathbf{B}_{0}\right)
\label{b1}
\end{equation}
with $q$ the particle charge. The left hand side describes inertia while the right hand side accounts for electric and magnetic forces. Equation (\ref{b1}) has two solutions,
\begin{equation}
\mathbf{v}_{0}=\Omega _{\pm }r\mathbf{e}_{\theta },  \label{b11}
\end{equation}
associated with the two roots $\Omega _{\pm }$ of the characteristic equation 
\begin{equation}
2\frac{\Omega _{\pm }}{\omega _{c}}=-\frac{q}{|q|}\left[1\pm \sqrt{1-4\frac{q}{|q|}\frac{\Omega }{\omega _{c}}}\right]
\end{equation}
obtained by substituting Eq. (\ref{b11}) into Eq. (\ref {b1})~\citep{Brillouin1945,Davidson2001}. Consistent with earlier notations, we use here $\omega_c = |q|B_0/m$ as the unsigned cyclotron frequency.

In the remaining of this study we consider only the small root associated with the slow $E\times B$ rotation, that is $\Omega_{-}\approx -\Omega $, as it is the rotation mode spontaneously observed quasi-neutral plasmas experiments. It also leads to $\Omega_{e}=\Omega_{i}$, or in other words solid body rotation at equal angular frequency for ions and electrons. However, since both slow $\Omega _{-}\approx -\Omega $ and fast $\Omega _{+}\approx -q\omega_{c}/|q|$ rotation modes have been observed and analysed in non-neutral plasmas~\citep{Davidson2001}, it is worth noting here that our findings could easily be extended and applied to higher angular velocities.

In the next two sections we will perturb this $\left[ \mathbf{v}_{0},\mathbf{E}_{0},\mathbf{B}_{0}\right] $ Brillouin equilibrium given by Eqs.~(\ref{b1}, \ref{b11}) with helical electrostatic and electromagnetic waves $\left[\mathbf{E}\left( r,\theta ,z,t\right) ,\mathbf{B}\left( r,\theta ,z,t\right)\right] $ with a space and time structure of the form 
\begin{equation}
\mathbf{A}=\mathbf{a}\left( r\right) \exp j\left( \omega t-l\theta -\beta
z\right).  \label{A0}
\end{equation}
According to Appendix~\ref{Sec:Appendix_A}, $\widehat{L}_{z}\mathbf{A}=-l\mathbf{A}$ and $\widehat{P}_{z}\mathbf{A}=-\beta \mathbf{A}$, so that one quantum of these waves carry an orbital angular momentum $l\hbar$ and a linear angular momentum $\beta\hbar$. We will then derive the expressions describing the rotating plasma response to this helical electromagnetic wave to obtain the dispersion relation. The dispersion relation will finally be used to identify Faraday-Fresnel rotation and Faraday-Fresnel splitting, that is to say the effect of plasma rotation on OAM-carrying waves. The following set of relations 
\begin{eqnarray}
\bm{\nabla }\times \left( r\mathbf{e}_{\theta }\times \mathbf{A}\right)
&=&jl\mathbf{A}+r\left( \bm{\nabla }\cdot \mathbf{A}\right) \mathbf{e}_{\theta },  \label{A1} \\
\bm{\nabla }\times \left( \mathbf{e}_{z}\times \mathbf{A}\right)
&=&j\beta \mathbf{A}+\left( \bm{\nabla }\cdot \mathbf{A}\right) \mathbf{e}_{z},  \label{A3} \\
\bm{\nabla }\cdot \left( r\mathbf{e}_{\theta }\times \mathbf{A}\right)
&=&2\mathbf{A\cdot e}_{z}-r\mathbf{e}_{\theta }\cdot \mathbf{\nabla \times A},  \label{A2} \\
\bm{\nabla }\cdot \left( \mathbf{e}_{z}\times \mathbf{A}\right) &=&-\mathbf{e}_{z}\cdot \mathbf{\nabla \times A},  \label{A4} \\
\left( \mathbf{A}\cdot \bm{\nabla }\right) r\mathbf{e}_{\theta } &=&\mathbf{e}_{z}\times \mathbf{A},  \label{A5} \\
\left( r\mathbf{e}_{\theta }\cdot \bm{\nabla }\right) \mathbf{A} &=&-jl\mathbf{A}+\mathbf{e}_{z}\times \mathbf{A},  \label{BA6}
\end{eqnarray}
will be particularly useful to solve Maxwell's equation for the TG and WH modes.

\section{Faraday-Fresnel rotation for Trivelpiece-Gould waves}
\label{Sec:IV}

In a magnetised plasma column close to the plasma and cyclotron frequencies both cyclotron and Langmuir oscillations are expected to propagate along the magnetic field. The analysis of these electrostatic elementary excitations suggests considering a potential electric field $\mathbf{E}$ such
that $\bm{\nabla }\times \mathbf{E}=\mathbf{0}$~\citep{Davidson2001,Stenzel2016}. This electric
field derives from a potential $\phi $ whose dynamics is associated with an oscillating electron space charge $-ne$. We therefore consider the propagation of potential and density perturbations 
\begin{eqnarray}
\phi &=&\phi \left( r\right) \exp j\left( \omega t-l\theta -\beta z\right),  \label{p4} \\
n &=&n\left( r\right) \exp j\left( \omega t-l\theta -\beta z\right),
\end{eqnarray}
over the equilibrium static radial electric potential $\phi_0 = -B_{0}\Omega r^{2}/2$ from which derives the electric field $\mathbf{E}_0$ driving the equilibrium Brillouin rotation and uniform densities $n_0$. Equation (\ref{b1}) provides the unperturbed momentum balance for a Brillouin flow while the relation $\bm{\nabla }\cdot n_{0}\mathbf{v}_{0}=0$ yields the unperturbed charge conservation relation. 

\subsection{Disperson relation}

For a weak electronic perturbation of this Brillouin flow the linearised momentum balance, Eq. (%
\ref{O1}), and the linearised charge conservation relation, Eq. (\ref{O2}), 
\begin{eqnarray}
j\omega \mathbf{v}+\left( \mathbf{v}_{0}\cdot \bm{\nabla }\right) 
\mathbf{v}+\left( \mathbf{v}\cdot \bm{\nabla }\right) \mathbf{v}_{0} &=&%
-\frac{e}{m_e}\left( \mathbf{E}+\mathbf{v}\times \mathbf{B}_{0}\right),
\label{O1} \\
j\omega n+n_{0}\bm{\nabla }\cdot \mathbf{v}+\mathbf{v}_{0}\cdot \mathbf{%
\nabla }n &=&0,  \label{O2}
\end{eqnarray}
yield a set of coupled equations relating $n$ to $\phi$. Making use of Eqs. (\ref{A5}, \ref{BA6}), Eq. (\ref{O1}) can be rearranged as 
\begin{equation}
j\left( \omega -l\Omega \right) \mathbf{v}+\left( 2\Omega -\omega
_{ce}\right) \mathbf{e}_{z}\times \mathbf{v}=\frac{e}{m_e}\bm{\nabla }%
\phi,  \label{O3}
\end{equation}
while Eq. (\ref{O2}) can be rewritten as 
\begin{equation}
n_{0}\bm{\nabla }\cdot \mathbf{v}=-j\left( \omega -l\Omega \right) n%
.  \label{O4}
\end{equation}
We then apply the three linear operators $\bm{\nabla }\times$, $\bm{\nabla }\cdot $ and $\mathbf{e}_z\cdot$ to Eq.~ (\ref{O3}) to obtain three relations relating the divergence, the curl and the longitudinal component of the velocity $\mathbf{v}$. With the help of Eqs~(\ref{A3}, \ref{A4}) this yields
\begin{eqnarray}
j\left( \omega -l\Omega \right) \bm{\nabla }\times \mathbf{v}+\left(
2\Omega -\omega _{ce}\right) \left[ \left( \bm{\nabla }\cdot \mathbf{v}%
\right) \mathbf{e}_{z}+j\beta \mathbf{v}\right] &=&\mathbf{0},
\label{O5} \\
j\left( \omega -l\Omega \right) \bm{\nabla }\cdot \mathbf{v}-\left(
2\Omega -\omega _{ce}\right) \mathbf{e}_{z}\cdot \bm{\nabla }\times 
\mathbf{v} &=&{\omega_{pe}}^{2}\frac{n}{n_{0}},  \label{O6} \\
\left( \omega -l\Omega \right) \mathbf{e}_{z}\cdot \mathbf{v} &=&-\frac{e}{m_e}%
\beta \phi,  \label{O7}
\end{eqnarray}
where we have used Poisson equation to eliminate $\Delta \phi $ in Eq. (\ref{O6}). Velocity dependent terms $\bm{\nabla }\times \mathbf{v}$, $\bm{\nabla }\cdot\mathbf{v}$ and $\mathbf{v}$ are then eliminated between Eqs. (\ref{O4}, \ref{O5}, \ref{O6}, \ref{O7}) to write the oscillating space charge $-ne$ as a function of the oscillating potential $\phi$
\begin{equation}
-\frac{ne}{\varepsilon _{0}}=\beta ^{2}\frac{{\omega_{pe}}^{2}}{\left( \omega
-l\Omega \right) ^{2}}\frac{\left(\omega _{ce}-2\Omega\right) ^{2}}{\omega
_{pe}^{2}+\left(\omega _{ce}-2\Omega\right) ^{2}-\left( \omega -l\Omega
\right) ^{2}}\phi.
\end{equation}
Poisson equation $\Delta \phi $ = $ne/\varepsilon _{0}$ for the perturbation potential $\phi \left( r\right) $ given in Eq.~(\ref{p4}) finally leads to the final equation describing TG modes propagation 
\begin{equation}
\frac{1}{r}\frac{\partial }{\partial r}r\frac{\partial \phi }{\partial r}-\frac{l^{2}}{r^{2}}\phi +\beta ^{2}T_{l}^{2}\phi =0  \label{O8}
\end{equation}
where we have defined the parameter
\begin{equation}
T_{l}^{2}=\frac{{\omega_{pe}}^{2}\left(\omega _{ce}-2\Omega\right) ^{2}}{\left( \omega -l\Omega \right) ^{2}\left[ {\omega_{pe}}^{2}+\left(\omega _{ce}-2\Omega\right) ^{2}-\left( \omega -l\Omega \right) ^{2}\right] }-1.  \label{O11}
\end{equation}
Looking at Eqs.~(\ref{O8}) and (\ref{O11}), one first verifies that the classical TG relation Eq.~(\ref{dt}) first derived by \cite{Trivelpiece1959} is recovered in the limit of zero rotation $\Omega=0$. One additionally notes that $T_{+l}=T_{-l}$ in this same limit of zero rotation, but that $T_{+l}\neq T_{-l}$ for $\Omega \neq 0$. Finally, a closer examination of Eq.~(\ref{O11}) shows that the $\Omega $ dependent terms $\left(\omega _{ce}-2\Omega\right) $ can be traced back to the Coriolis effect in the plasma rest frame~\citep{Lehnert1962,Uberoi1970}, while the $\left(\omega -l\Omega \right) $ terms are to be attributed to the Doppler effect in the plasma rest frame.

\subsection{Solutions as superpositions of eigenmodes with opposite OAM contents}

A general non-singular solution of Poisson equation Eq. (\ref{O8}) describing TG modes with an OAM content such that $\left| L_{z}/P_{z}\right| =l/\beta$ consists in a linear combination of two counter-rotating Bessel modes 
\begin{equation}
\phi =\phi _{+}J_{l}\left( \beta T_{l}r\right) \exp j\left( \omega t-l\theta
-\beta z\right) +\phi _{-}J_{-l}\left( \beta T_{-l}r\right) \exp j\left(
\omega t+l\theta -\beta z\right),  \label{O10}
\end{equation}
where $l\in \Bbb{N}$ is a positive number and the mode amplitudes $\phi _{\pm }$ are assumed to be set by the antenna. Classically TG modes are studied as confined modes in a conducting cylindrical cavity of length $L$ and radius $a$, which leads to the boundary conditions $E_{\theta}\left( a\right) = E_{z}\left( a\right) =0$, and from there to $J_{l}\left( \beta T_{l}a\right) =0$. The axial wave vector $\beta $ is then determined through the usual requirement of standing wave along $z$, that is $\beta L=p\pi $ with $p\in \Bbb{Z}$. This set of conditions fully determines the
resonant mode frequency in the cylindrical cavity through $J_{l}\left( p\pi T_{l}a/l\right) =0$. 

Instead of following this approach, we look here for free TG modes propagating along $z$. Specifically we consider free TG modes displaying the same radial structure $J_{l}\left( \alpha r\right) $. In other words, instead of the solution in Eq.~(\ref{O10}) consisting in the superposition of modes with opposite azimuthal mode numbers $\pm l$, different radial structure and identical axial wave vectors $\beta$, we consider non-singular solutions
\begin{equation}
\phi =\phi _{0}J_{l}\left( \alpha r\right) \exp j\left( \omega t-l\theta-\beta _{l}z\right) +\phi _{0}J_{-l}\left( \alpha r\right) \exp j\left(\omega t+l\theta -\beta _{-l}z\right).  \label{O28}
\end{equation}
that are the superposition of modes with opposite azimuthal mode numbers $\pm l$, identical radial structure but different axial wave vectors  $\beta _{+l}\neq \beta _{-l}$. Both the frequency $\omega$ and the transverse parameter $\alpha$ are set by the antenna, whereas the wave vectors $\beta _{+l}$ and $\beta _{-l}$ depend on the rotating plasma properties through the condition 
\begin{equation}
\beta _{\pm l}T_{\pm l}(\omega ,\omega _{ce},\omega _{pe},\Omega )=\alpha 
 \label{DR1}
\end{equation}
which guarantees that $\phi$ verifies the radial Poisson equation Eq.~(\ref{O8}). Following the analysis given in Appendix~\ref{Sec:Appendix_A}, the longitudinal electric field associated with the $\pm l$ modes in Eq.~(\ref{O28}) are eigenvectors of $S_{z}$ with $0$ eigenvalue and of $L_{z}$ with eigenvalues $\pm l$, and thus also of the sum $L_{z}+S_{z}$.


\subsection{Faraday-Fresnel rotation}

Going back to Eq.~(\ref{dtg}), the normalised difference between these wave vectors $\beta _{+l}$ and $\beta_{-l}$ is 
\begin{equation}
\frac{\beta _{+l}-\beta _{-l}}{\beta _{+l}+\beta _{-l}}=\frac{T_{-l}-T_{+l}}{T_{-l}+T_{+l}} =\frac{\Omega}{2\left.
T_{l}\right| _{\Omega =0}} \left.\frac{\partial \left(T_{-l}-T_{+l}\right) }{\partial \Omega}\right| _{\Omega =0} + \mathcal{O}\left(\frac{\Omega}{\omega} \right) ^{2}.
\end{equation}
From Eq.~(\ref{O11}), the lowest order in $\Omega/\omega$ estimate for this phase shift then writes
\begin{equation}
\left. \frac{\beta _{+l}-\beta _{-l}}{\beta _{+l}+\beta _{-l}}\right|
_{TG}=\frac{\Omega}{4T_{l}^{2}}\left( \left. \frac{\partial T_{-l}^{2}}{\partial \Omega}\right| _{\Omega =0} -\left. \frac{\partial T_{+l}^{2}}{\partial \Omega}\right| _{\Omega =0}\right)  =-l\frac{\Omega 
}{\omega }F_{TG}\left( \omega _{pe},\omega _{ce},\omega \right),
\label{O14}
\end{equation}
where we have defined the form factor 
\begin{equation}
F_{TG}=\frac{{\omega_{pe}}^{2}}{\left( {\omega_{pe}}^{2}-\omega ^{2}\right) }%
\frac{{\omega_{ce}}^{2}}{\left( {\omega_{ce}}^{2}-\omega ^{2}\right) }\frac{%
\left( {\omega_{uh}}^{2}-2\omega ^{2}\right) }{\left( {\omega_{uh}}^{2}-\omega
^{2}\right) }  \label{ff1}
\end{equation}
which is $\mathcal{O}(1)$ at low frequency. 

Eq.~(\ref{O14}) shows that, indeed, plasma rotation introduces a phase shift between OAM-carrying TG eigenmodes with opposite azimuthal mode numbers, or in other words to Faraday-Fresnel rotation. The wave transverse phase structure is hence rotated as the wave propagates along the rotation and magnetic axis $\mathbf{e}_z$. Comparing now Eq.~(\ref{O14}) with the phase shift introduced by rotation between circularly polarised eigenmodes~\citep{Gueroult2019a,Gueroult2020}, that is to say to the SAM mechanical Faraday effect, one finds that the phase shifts are in both case proportional to the ratio of plasma angular frequency to wave angular frequency $\Omega/\omega$, but that in addition Eq.~(\ref{O14})  depends linearly on the azimuthal wave number $l$. This $\Omega/\omega$ scaling is also consistent with image rotation in an isotropic dielectric~\citep{Goette2007}.


\subsection{Possible antenna design}

To conclude this section we propose here a coupling structure designed to launch the mode Eq.~(\ref{O28}), \emph{i.~e} two TG modes with identical radial structure but opposite azimuthal mode numbers. For this we draw on the inductive antenna design reviewed by \cite{Stenzel2019}. 

\begin{figure}
\begin{center}
\includegraphics[width = 12cm]{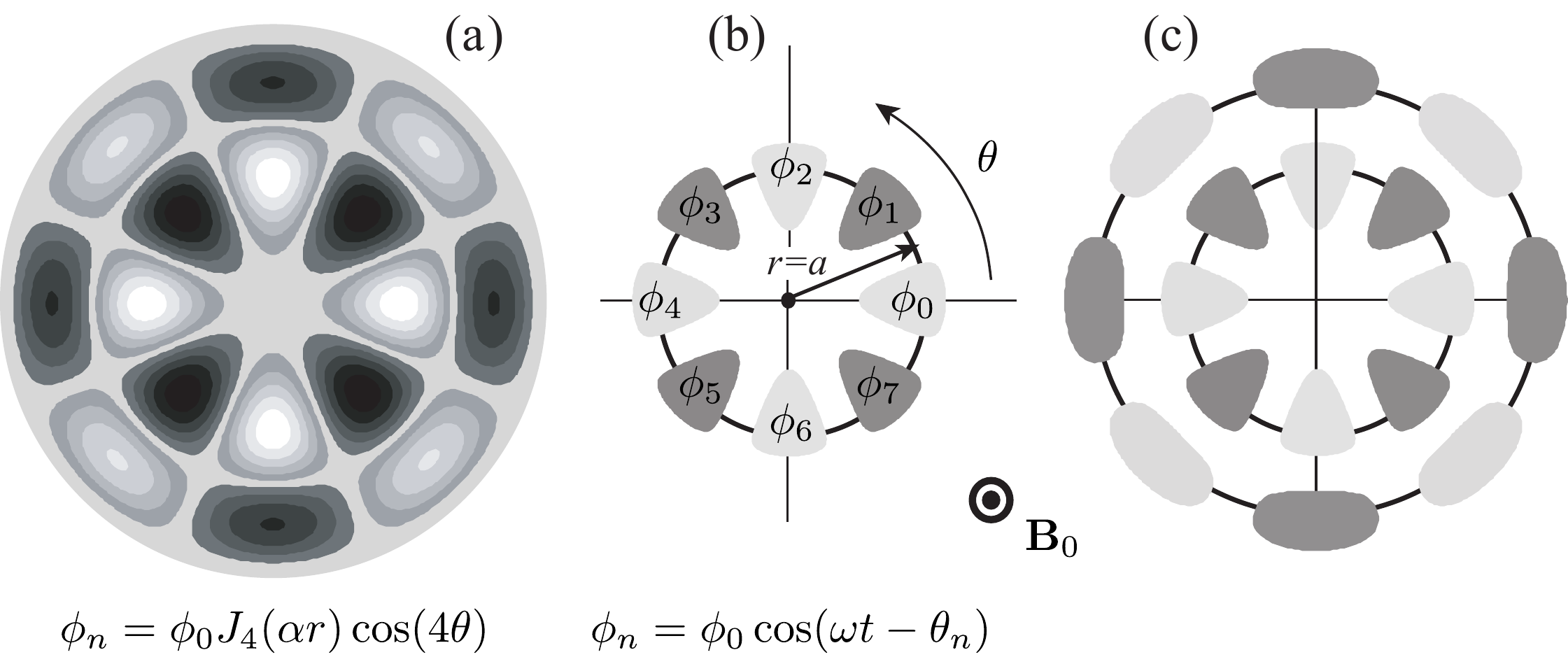}
\caption{(a) Bessel TG mode $l=4$ at $z = 0$ and $t = 0$. The radius $a$ of extrema on the first ring verifies $\alpha a = 5.317$. (b) Corresponding capacitive antenna with one ring of phased polarised plates located at $z=0$ and $r=a$. (c) Advanced antenna using two rings of polarised plates.}
\label{Fig:Fig1}
\end{center}
\end{figure}

The proposed configuration is shown in Fig.~\ref{Fig:Fig1}(b). It consists in a capacitive antenna made of $2l$ phased conductive plates positioned perpendicularly to the magnetic field $\mathbf{B}_{0}$ in the plane $z=0$ and arranged along the polar angle $\theta$ at 
\begin{equation}
\theta _{n}=n\frac{\pi}{l}, \qquad n\in \left[ 0,1,2,3...2l-1\right], 
\end{equation}
on a ring of radius $a$. The radius of the antenna ring $a$ is chosen to match the first extremum of $J_{l}\left( \alpha r\right)$. To both have a good impedance matching to the $\pm l$ TG modes components of Eq.~(\ref{O28}) and minimise the excitation of $\pm l^{\prime}$ satellites, the size and shape of the $2l$ plates are chosen similar to the size and shape of the pattern $\left| J_{l}\left( \alpha r\right) \cos \left( l\theta \right) \right| $ illustrated in Fig.~\ref{Fig:Fig1}(a) for $l=4$. The applied oscillating bias $\phi_n (t)$ on each of these azimuthally distributed $2l$ plates must then verify 
\begin{equation}
\phi_{n}\left( t\right) =\phi_{0}J_{l}\left( \alpha a\right) \left[ \exp -jn\pi +\left( -1\right)^{l}\exp jn\pi \right] \exp j\omega t, \qquad n\in \left[ 0,1,2,3...2l-1\right],   \label{O9}
\end{equation}
to preferentially excite the targeted modes 
\begin{equation}
\phi \left( r,\theta ,z,t\right) =\phi _{0}\left[ J_{+l}\left( \alpha r\right) \exp j\left( -l\theta -\beta _{+l}z\right) +J_{-l}\left( \alpha r\right) \exp j\left( l\theta -\beta _{-l}z\right) \right] \exp j\omega t.  \label{O21}
\end{equation}
Going further, an improved antenna design could be constructed by adding a second ring of $2l$ phased conductive plates near the second extremum of $J_{\left| l\right| }\left( \alpha r\right)$, as illustrated in Fig.~\ref{Fig:Fig1}(c).

\section{Faraday-Fresnel rotation for Whistler-Helicon waves}
\label{Sec:V}

Whistler waves in a cold magnetised plasma at rest are right circularly polarised waves whose propagation below $\omega _{pe}$ is sustained by the electron Hall current. Because of the low-frequency nature of these waves,
the displacement current in Maxwell-Amp\`{e}re equation can be neglected, and propagation is characterised by the dispersion relation given in Eq.~(\ref{dw}). The antennae design to excite this WH branch~\citep{Stenzel2014,Stenzel2019,Urrutia2015}, as well as the properties associated with their angular momentum dynamics, have been extensively investigated in Refs.~\citep{Urrutia2015a,Urrutia2016,Stenzel2015,Stenzel2015a,Stenzel2018}. 

Consider Whistler-Helicon (WH) electric $\left( \mathbf{E}\right) $ and magnetic $\left( \mathbf{B}\right) $ fields with helical phase fronts propagating along the $z$ axis and rotating around this axis
\begin{eqnarray}
\mathbf{E} &=&\mathbf{E}\left( r\right) \exp j\left( \omega t-l\theta -\beta
z\right), \\
\mathbf{B} &=&\mathbf{B}\left( r\right) \exp j\left( \omega t-l\theta -\beta
z\right).
\end{eqnarray}
This electromagnetic perturbation is solution to Maxwell-Faraday and Maxwell-Amp\`{e}re equations 
\begin{eqnarray}
\bm{\nabla }\times \mathbf{E} &=&-j\omega \mathbf{B},  \label{C1}
\\
\bm{\nabla }\times \mathbf{B} &=&\mu _{0}\mathbf{J},  \label{C2}
\end{eqnarray}
and is associated with an electron velocity 
\begin{equation}
\mathbf{v}=\mathbf{v}\left( r\right) \exp j\left( \omega t-l\theta -\beta
z\right)
\end{equation}
and an electron current 
\begin{equation}
\mathbf{J}=-n_{0}e\mathbf{v}-ne\mathbf{v}_{0},
\end{equation}
with $\mathbf{v}_{0}$ the equilibrium Brillouin velocity given in~Eq.~(\ref{b11}). However, Maxwell-Amp\`{e}re equation Eq.~(\ref{C2}) can be used to get $\bm{\nabla }\cdot \mathbf{J}=0$ and from there $j\omega ne=0$, so that the current simply writes $\mathbf{J}=-n_{0}e\mathbf{v}$. 

\subsection{Dispersion relation}

The linearised momentum balance 
\begin{equation}
j\omega \mathbf{v}+\left( \mathbf{v}_{0}\cdot \bm{\nabla }\right) 
\mathbf{v}+\left( \mathbf{v}\cdot \bm{\nabla }\right) \mathbf{v}_{0}=-\frac{e}{m_e}\left( \mathbf{E}+\mathbf{v}_{0}\times \mathbf{B}+\mathbf{v}\times \mathbf{B}_{0}\right)
\end{equation}
allows, with the help of Eqs. (\ref{A5}, \ref{BA6}), to write the electron current as a function of the fields
\begin{equation}
j\left( \omega -l\Omega \right) \mathbf{J}+\left( 2\Omega -\omega
_{ce}\right) \mathbf{e}_{z}\times \mathbf{J}={\omega_{pe}}^{2}\varepsilon
_{0}\left( \mathbf{E}+\Omega r\mathbf{e}_{\theta }\times \mathbf{B}\right) 
.  \label{C3}
\end{equation}
The operator $\bm{\nabla }\times$ is then applied to Eq.~(\ref{C3}) to eliminate both the electric field $\mathbf{E}$ and the current $\mathbf{J}$. Using Eqs. (\ref{A1}, \ref{A3}, \ref{C1}, \ref{C2}) finally yields the wave equation 
\begin{equation}
\bm{\nabla }\times \bm{\nabla }\times \mathbf{B}-\beta \frac{\omega _{ce}-2\Omega
}{\omega -l\Omega }\bm{\nabla }\times \mathbf{B}+\frac{%
{\omega_{pe}}^{2}}{c^{2}}\mathbf{B=0}.  \label{C5}
\end{equation}
One verifies that Eq.~(\ref{C5}) reduces to the classical WH equation~\citep{Klozenberg1965} for $\Omega=0$. In addition, Eq.~(\ref{C5}) can be rewritten as 
\begin{equation}
\left( \beta _{1}-\bm{\nabla }\times \right) \left( \beta _{2}-
\bm{\nabla} \times \right) \mathbf{B}=\mathbf{0}  \label{C4}
\end{equation}
with
\begin{eqnarray}
\beta_{1}\beta_{2} &=&\frac{{\omega_{pe}}^{2}}{c^{2}}, \\
\beta_{1}+\beta_{2} &=&\beta \frac{\omega_{ce}-2\Omega}{\omega -l\Omega 
}.
\end{eqnarray}
The associated second order algebraic equation $x^{2}-(\beta_{1}+\beta _{2}) x+\beta_{1}\beta_{2} = 0$ has two real roots. The large root is related to TG modes whereas the small one describes the dispersive properties of the WH branch 
\begin{equation}
\beta _{W}=\frac{\beta }{2}\frac{\omega _{ce}-2\Omega}{\omega -l\Omega }\left[ 1-\sqrt{1-4\frac{{\omega_{pe}}^{2}}{{\beta _{W}}^{2}c^{2}}\left( \frac{\omega -l\Omega }{\omega _{ce}-2\Omega}\right) ^{2}}\right] \approx \frac{{\omega_{pe}}^{2}}{\beta _{W} c^{2}}\frac{\omega -l\Omega }{\omega _{ce}-2\Omega },
\end{equation}
where we have used the WH ordering $\omega _{ce}\gg \omega$ to obtain an estimate for $\beta _{W}$ in the last equation. Note that we recover Eq.~(\ref{dw}) for $\Omega =0$, and also that $\beta _{W}\left( +l\right) \neq \beta _{W}\left(-l\right) $ for $\Omega \neq 0$. Similarly to the analysis done above for TG modes, further examination shows that the $\left(\omega_{ce}-2\Omega\right) $ terms are associated with the Coriolis effect in the plasma rest frame~\citep{Lehnert1962,Uberoi1970} while the $\left( \omega -l\Omega \right) $ terms come from the Doppler effect in the plasma rest frame.

Now by introducing
\begin{equation}
K_{l}=\frac{{\omega_{pe}}^{2}}{c^{2}}\frac{\omega -l\Omega }{\omega
_{ce}-2\Omega }  \label{k5}
\end{equation}
such that $K_{+l}=K_{-l}$ for $\Omega =0$ but $K_{+l}\neq K_{-l}$ for $\Omega \neq 0$ and
\begin{equation}
K_{0}=\frac{{\omega_{pe}}^{2}}{c^{2}}\frac{\omega }{\omega _{ce}},
\end{equation}
one finds that Maxwell-Ampere equation for the WH branch reduces to 
\begin{equation}
\bm{\nabla }\times \mathbf{B} = \left( K_{l}/\beta \right) \mathbf{B}.
\end{equation}
Applying the operator $\bm{\nabla }\times $ to this last relation leads to the wave
equation
\begin{equation}
\Delta \mathbf{B}+\left( K_{l}^{2}/\beta ^{2}\right) \mathbf{B} =\mathbf{0}
\end{equation}
whose $z$ component writes 
\begin{equation}
\frac{1}{r}\frac{\partial }{\partial r}r\frac{\partial B_{z}}{\partial r}-%
\frac{l^{2}}{r^{2}}B_{z}+\left( \frac{K_{l}^{2}}{\beta ^{2}}-\beta
^{2}\right) B_{z}=0.  \label{C6}
\end{equation}

\subsection{Solutions as superpositions of eigenmodes with opposite OAM contents}

Following a similar approach to that used for TG modes,  we look for solutions of Eq.~(\ref{C6}) of the form 
\begin{equation}
B_{z}=B_{z0}\left[ J_{l}\left( \alpha r\right) \exp j\left( \omega t-l\theta
-\beta _{l}z\right) +J_{-l}\left( \alpha r\right) \exp j\left( \omega
t+l\theta -\beta _{-l}z\right) \right],  \label{c17}
\end{equation}
that is to say superpositions of eigenmodes with opposite azimuthal mode numbers $\pm l$, identical radial structure $J_{l}\left( \alpha r\right) $ but different axial wave vectors  $\beta _{+l}\neq \beta _{-l}$. Both the frequency $\omega $ and the transverse parameter $\alpha $ are set by the antenna while the wave vectors $\beta _{\pm l}$ is determined by the properties of the rotating plasma through the relation 
\begin{equation}
\beta _{\pm l}^{-2}K_{\pm l}^{2}(\omega ,\omega _{ce},\omega _{pe},\Omega
)-\beta _{\pm l}^{2}=\alpha ^{2}  \label{DR2}
\end{equation}
which guarantees that $B_z$ verifies Eq. (\ref{C6}). Separating the $\pm l$ contributions, Eq.~(\ref{DR2}) can be rewritten as 
\begin{equation}
\beta _{l}^{-2}K_{l}^{2}-\beta _{l}^{2}=\beta _{-l}^{-2}K_{-l}^{2}-\beta
_{-l}^{2}=\alpha ^{2}  \label{D23}
\end{equation}
whose solutions are
\begin{equation}
\beta _{\pm l}^{2}=\sqrt{\alpha ^{4}+4K_{\pm l}^{2}}-\alpha ^{2}.
\label{ap2}
\end{equation}

\subsection{Faraday-Fresnel rotation}

Expanding $K_{\pm l}^{2}$ given in Eq. (\ref{k5}) for the small parameter $\Omega/\omega$ and assuming WH ordering $\omega _{ce}\gg \omega$ yields 
\begin{equation}
K_{\pm l}^{2}=K_{0}^{2}\left( 1\mp 2l\frac{\Omega }{\omega}\right) + \mathcal{O}\left[\left(\frac{\Omega }{\omega }\right)^2\right].
\label{Eq:Kll}
\end{equation}
Meanwhile, from Eq.~(\ref{ap2}), 
\begin{equation}
\frac{\beta _{+l}^{2}-\beta _{-l}^{2}}{\left( \beta _{+l}+\beta _{-l}\right)
^{2}} = \frac{\sqrt{\alpha ^{4}+4K_{+l}^{2}}-\sqrt{\alpha
^{4}+4K_{-l}^{2}}}{2\sqrt{\alpha ^{4}+4K_{0}^{2}}-2\alpha ^{2}}.
\end{equation}
Plugging Eq~(\ref{Eq:Kll}) into this last expression finally leads to the lowest order in $\Omega/\omega$ estimate for the OAM phase shift
\begin{equation}
\left. \frac{\beta _{+l}-\beta _{-l}}{\beta _{+l}+\beta _{-l}}\right|
_{WH}=-l\frac{\Omega }{\omega }F_{WH}\left( \omega _{pe},\omega _{ce},\omega
\right),  \label{o45}
\end{equation}
where we have defined the form factor
\begin{equation}
F_{WH}=\frac{4K_{0}^{2}}{\alpha ^{4}+4K_{0}^{2}-\alpha ^{2}\sqrt{\alpha
^{4}+4K_{0}^{2}}}.  \label{ff2}
\end{equation}
Similarly to Eq.~(\ref{ff1}) which was obtained for TG modes, Eq.~(\ref{ff2}) shows that plasma rotation introduces a phase shift between OAM-carrying WH eigenmodes with opposite azimuthal mode numbers, or in other words to Faraday-Fresnel rotation. In both instances, the OAM phase shift is proportional to the plasma angular frequency to wave angular frequency ratio $\Omega/\omega$, and to the azimuthal wave number $l$. However, closer examination of Eq.~(\ref{ff1}) and Eq.~(\ref{ff2}) shows that the form factor $F$ characterising Faraday-Fresnel rotation does depend on $\alpha$ (\emph{i.~e.} the wave radial structure) for WH modes, while it did not for TG modes.

\subsection{Possible antenna design}

To conclude this section we point here to a coupling structure designed to launch WH modes of the type Eq.~(\ref{c17}). This type of antenna has been extensively investigated, both theoretically and experimentally, in Refs.~\citep{Stenzel2014,Urrutia2015}, and we refer the readers to the review paper by~\cite{Stenzel2019} for an in depth analysis of their properties.

\begin{figure}
\begin{center}
\includegraphics[width = 12cm]{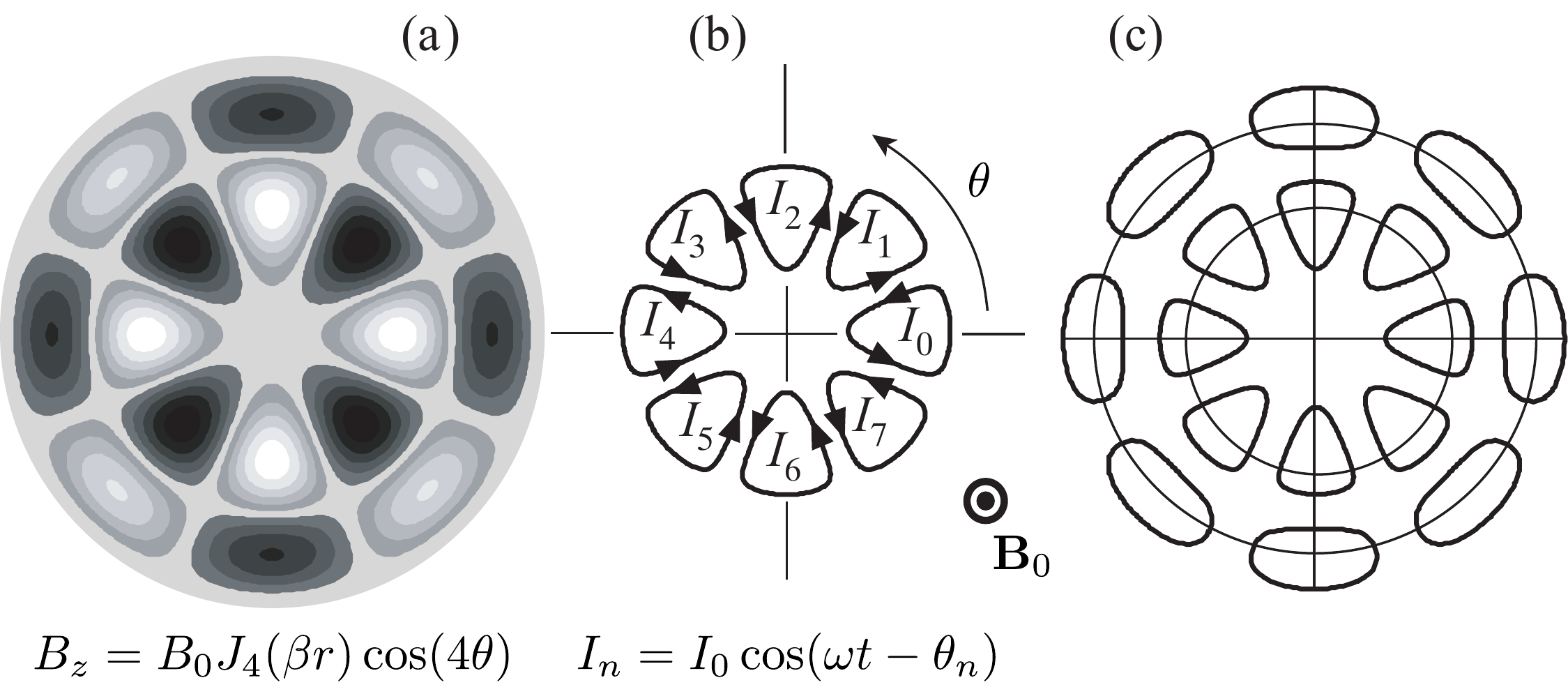}
\caption{Bessel WH mode $l=4$ at $z = 0$ and $t = 0$. The radius $a$ of extrema on the first ring verifies $\alpha a = 5.317$. (b) Corresponding inductive antenna with one ring of phased current loops located at $z=0$ and $r=a$. (c) Advanced antenna using two rings of current loops.}
 \label{Fig:Fig2}
\end{center}
\end{figure}

The envisioned configuration is shown in Fig.~\ref{Fig:Fig2}(b). It consists in an inductive antenna made of $2l$ phased current loops positioned perpendicularly to the magnetic field $\mathbf{B}_{0}$ in the plane $z=0$ and arranged along the polar angle $\theta$ at 
\begin{equation}
\theta_{n}=n\pi /l,\qquad n\in \left[ 0,1,2,3...2l-1\right],
\end{equation}
on a ring of radius $a$. The currents $I_{n}\left( t\right)$ of these $2l$ current loops are phased according to 
\begin{equation}
I_{n}\left( t\right) =B_{z0}J_{l}\left( \alpha a\right) \left[ \exp -jn\pi
+\left( -1\right) ^{l}\exp +jn\pi \right] \exp j\omega t,\qquad n\in \left[ 0,1,2,3...2l-1\right],  \label{IN}
\end{equation}
while the radius of the antenna ring $a$ is chosen to match the first extremum of $J_{l}\left(\alpha r\right) $. The set of parameters $(\alpha,B_{z0}, l, \omega)$ are hence all set by the antenna array. 

To both have a good impedance matching to the $\pm l$ WH modes components of Eq.~(\ref{c17}) and minimize the excitation of $\pm l^{\prime }$ satellites the size and shape of the $2l$ loops should be taken as close as possible to the size and shape of the wave pattern $\left| J_{l}\left( \alpha r\right)\cos \left( l\theta \right) \right| $ illustrated in Fig.~\ref{Fig:Fig2}(a). Alternatively, an improved antenna array can be designed with a second ring of $2l$ current loops near the second extremum of $J_{\left| l\right| }\left( \alpha r\right) $, as illustrated in Fig.~\ref{Fig:Fig2}(c). 

\section{Faraday-Fresnel splitting in a rotating plasma}
\label{Sec:VI}

Consider now a wave packet in a plasma exhibiting both a transverse structure and a longitudinal structure. 
The rotating phase of the wave packet travels at the phase velocity $\omega /\beta $ along
the $z$ axis, whereas the longitudinal envelope of the wave packet travels at the group velocity $\partial \omega /\partial \beta$ along this axis~\citep{Stix1992,Urrutia2015}.

It has been shown in Sections~\ref{Sec:IV} and \ref{Sec:V} that plasma rotation introduces a phase shift between TG and WH OAM-carrying modes with opposite azimuthal mode numbers, and that for sufficiently small plasma angular frequency $\Omega$ the longitudinal wave vectors of these rotating modes write
\begin{equation}
\beta _{\pm l}=\left. \beta \right| _{\Omega =0}\left[ 1\mp l\frac{\Omega }{\omega }F\left( \omega _{pe},\omega _{ce},\omega \right) \right].
\label{beta}
\end{equation}
In these conditions the longitudinal phase velocity of these modes then write
\begin{equation}
V_{\varphi ,\pm l} = \omega /\beta _{\pm l} = V_{\varphi }\left[ 1\pm l\frac{\Omega }{\omega }F\right]  \label{o1236}
\end{equation} 
with $V_{\varphi }$ = $\left. \omega /\beta \right| _{\Omega =0}$ the phase velocity along the $z$ axis in a plasma at rest $\Omega =0$. Meanwhile, deriving Eq.~(\ref{beta}) with respect to $\omega$ yields
\begin{equation}
V_{g,\pm l}^{-1}=V_{g}^{-1}\left[ \left( 1\mp l\frac{\Omega }{\omega }F\right) \mp \frac{V_{g}}{V_{\varphi }}l\Omega \left( \frac{\partial F}{\partial \omega }-\frac{F}{\omega }\right) \right]
\end{equation}
with $V_{g,\pm l} = \partial \omega /\partial \beta_{\pm l}$ the longitudinal group velocity of these $\pm l$ modes, and $V_g = \partial \left.\beta\right|_{\Omega =0}/\partial\omega$ the group velocity for zero rotation. Assuming that all factors involving the small parameter $\Omega /\omega$ are small, inverting this last relation yields the result 
\begin{equation}
\frac{V_{g,\pm l}}{V_{g}}=1\pm l\frac{\Omega }{\omega }\left[ \left( 1-\frac{V_{g}}{V_{\varphi }}\right) F+\frac{V_{g}}{V_{\varphi }}\omega \frac{\partial F}{\partial \omega }\right].  \label{o46}
\end{equation}

Eq.~(\ref{o46}) shows that in the presence of rotation ($\Omega\neq 0$) the longitudinal group velocity of OAM-carrying eigenmodes with opposite azimuthal wave numbers - such as the TG and WH modes studied in Section~\ref{Sec:IV} and \ref{Sec:V}  - is different, whereas they are equal absent rotation $\Omega=0$. A single envelope wave packet consisting of a mix of these OAM-carrying eigenmodes with opposite azimuthal wave numbers will hence be split into two envelopes. By analogy with the Faraday splitting of circularly polarised eigenmodes (SAM)~\citep{Weng2017}, we refer to this new plasma effect as Faraday-Fresnel splitting.

\section{Conclusions}
\label{Sec:Conclusion}

In this study we have identified and described two new plasma wave effects which arise as a result of plasma rotation. These effects are traced back to the adiabatic coupling which takes place between the orbital angular momentum (OAM) of a rotating wave and the OAM of the rotating plasma. 

Through an analysis of the response of a rotating magnetised plasma to perturbations carrying OAM, it is shown that plasma rotation introduces a phase shift between OAM-carrying eigenmodes having opposite azimuthal modes numbers. As a result the transverse wave pattern of a wave obtained by superposition of these counter-rotating OAM eigenmodes rotates as the wave propagates. Because of the analogy both with the classical Faraday effect in a magnetised plasma - which relates to wave's spin angular momentum (SAM) - and the rotatory photon drag known in rotating isotropic dielectric - which relates to the wave's OAM, this plasma effect is referred to as Faraday-Fresnel rotation. The dependence of this phase shift on plasma parameters is derived analytically here for two classical plasma modes, namely, Trivelpiece-Gould (TG) and Whistler-Helicon (WH) modes, as shown in Eq.~(\ref{O14}) and Eq.~({\ref{o45}), respectively.

The difference in longitudinal wave vectors for counter-rotating OAM eigenmodes is further shown to translate to different longitudinal group velocity for each of these modes, as given in Eq.~(\ref{o46}). A direct consequence of this result predicted here is the splitting of a wave packet containing a superposition of these modes, that is to say that the two counter-rotating eigenmodes will be progressively spatially separated. By analogy with the Faraday splitting of circularly polarised eigenmodes (SAM), this new plasma effect is referred to as Faraday-Fresnel splitting. 

Searching for the basic physic mechanism behind these effects, a closer examination of Eq.~(\ref{O11}) and (\ref{k5}) reveals that rotation manifests itself in the wave propagation properties through two additional terms besides the usual $\omega _{pe}/\omega$ and $\omega _{ce}/\omega$ terms associated with dispersion. The first one is $\left(\omega _{ce}- 2\Omega\right)$ and simply accounts for the modification of the magnetic force by the
Coriolis force as the rest frame of the plasma is rotating at $\Omega$. The second is $\left( \omega -l\Omega \right)$ and is a consequences of the rotational Doppler shift as the rest frame of the plasma is rotating at $\Omega$. One is however typically used to find centrifugal forces affecting the dynamics alongside Coriolis forces and Doppler shifts in rotating media. One might thus be left questioning what the role of centrifugal forces is here. The answer is not to be found in Eq.~(\ref{O11}) and (\ref{k5}) but rather in the transverse structure of the wave. Indeed, since 
\begin{equation}
J_{l}\left( \alpha r\right) \sim \frac{\left(\alpha r\right) ^{l}}{l!}+...,
\end{equation}
the maxima of the amplitude $J_{l}\left(\alpha r\right) $ are pushed radially outward for higher angular momentum $l$. This behaviour is reminiscent of the effect of centrifugal forces. For very high $l$ cylindrical waves, this pronounced radial outward localisation of the field amplitude is the so called whispering gallery effect associated with centrifugal forces.

Looking ahead, the Faraday-Fresnel rotation (FFR) and Faraday-Fresnel splitting (FFS) effects identified in this study open interesting perspectives. An immediate application would be to use FFR to measure rotation in laboratory experiments, and notably in tokamaks where rotation provides a means to control of turbulence and enhance
confinement. But these results also allows for the search of more exotic coupling mechanisms between the plasma wave's OAM and plasma rotation. To name a few, SAM cross polarisation scattering described by \cite{Lehner1989} might be extended to an OAM scheme where the vorticity fluctuation might now be probed through OAM-carrying waves. Another promising direction may lie the Zeeman splitting of the plasmon oscillation identified
by \cite{Robiche2008}, for which the effect of an additional rotation with OAM eigenmodes rather than with simple SAM eigenmodes should be evaluated. Besides these effects, another important question that remains to be examined is the complementary active or resonant part of the coupling between rotating OAM-carrying plasma waves and and the rotating plasma OAM.

Finally, a strong motivation to focus our analysis on TG and WH modes is that they are popular low frequency modes for plasma experiments. This should in turn facilitate experimental efforts to test these findings. Using a pair of antennae such as those described in Sections~\ref{Sec:IV} and \ref{Sec:V} in an emitter-receiver configuration, the results obtained here predict that the signal on the receiver should exhibit a double maximum profile as the receiver is rotated along the azimuthal angle, exposing that two $\pm l$ excited rotating waves are differentially dragged by the rotating plasma. It should however be stressed that the new results uncovered in this study for TG and WH modes are expected to be similarly found for other branches of plasma waves. One example is compressional rotating Alfv{\'e}n waves for which a relation akin to that obtained in Eq.~(\ref{C6}) can be derived from ideal MHD for the compressional magnetic field, suggesting that FFR will be at play. An in-depth systematic analysis of this broader problem is left for future studies.


\bibliographystyle{jpp}

\bibliography{JMROAM}

\begin{thebibliography}{73}
\expandafter\ifx\csname natexlab\endcsname\relax\def\natexlab#1{#1}\fi
\def\au#1{#1} \def\ed#1{#1} \def\yr#1{#1}\def\at#1{#1}\def\jt#1{\textit{#1}}
  \def\bt#1{#1}\def\bvol#1{\textbf{#1}} \def\vol#1{#1} \def\pg#1{#1}
  \def\publ#1{#1}\def\arxiv#1{#1}\def\org#1{#1}\def\st#1{\textit{#1}}

\bibitem[Allen {\em et~al.\/}(1992)Allen, Beijersbergen, Spreeuw \&
  Woerdman]{Allen1992}
{\sc \au{Allen, L.}, \au{Beijersbergen, M.~W.}, \au{Spreeuw, R. J.~C.} \&
  \au{Woerdman, J.~P.}} \yr{1992}  \at{Orbital angular momentum of light and
  the transformation of {L}aguerre-{G}aussian laser modes}.  \jt{Phys. Rev. A}
  \bvol{45}~(11),  \pg{8185--8189}.

\bibitem[Anderson {\em et~al.\/}(1958)Anderson, Baker, Bratenahl, Ise, Kunkel,
  Stone \& Furth]{Anderson1958}
{\sc \au{Anderson, O.~A.}, \au{Baker, W.~R.}, \au{Bratenahl, A.}, \au{Ise,
  J.~J.}, \au{Kunkel, W.~B.}, \au{Stone, J.~M.} \& \au{Furth, H.~P.}} \yr{1958}
   \bt{In {\em Proceedings of the United Nations international conference on
  the peaceful uses of atomic energy\/}},  \pg{pp. 155--160}.

\bibitem[Balbus \& Hawley(1998)]{Balbus1998}
{\sc \au{Balbus, S.~A.} \& \au{Hawley, J.~F.}} \yr{1998}  \at{Instability,
  turbulence, and enhanced transport in accretion disks}.  \jt{Rev. Modern
  Phys.}  \bvol{70}~(1),  \pg{1--53}.

\bibitem[Barnett \& Allen(1994)]{Barnett1994}
{\sc \au{Barnett, S.~M.} \& \au{Allen, L.}} \yr{1994}  \at{Orbital angular
  momentum and nonparaxial light beams}.  \jt{Opt. Commun.}  \bvol{110}~(5-6),
  \pg{670--678}.

\bibitem[Barnett {\em et~al.\/}(2017)Barnett, Babiker \& Padgett]{Barnett2017}
{\sc \au{Barnett, S.~M.}, \au{Babiker, M.} \& \au{Padgett, M.~J.}} \yr{2017}
  \at{Optical orbital angular momentum}.  \jt{Philos. Trans. R. Soc. A: Math.
  Phys. Eng. Sci.}  \bvol{375}~(2087),  \pg{20150444}.

\bibitem[Bekhtenev {\em et~al.\/}(1980)Bekhtenev, Volosov, Pal{'}chikov, Pekker
  \& Yudin]{Bekhtenev1980}
{\sc \au{Bekhtenev, A.~A.}, \au{Volosov, V.~I.}, \au{Pal{'}chikov, V.~E.},
  \au{Pekker, M.~S.} \& \au{Yudin, Y.~N.}} \yr{1980}  \at{Problems of a
  thermonuclear reactor with a rotating plasma}.  \jt{Nucl. Fusion}
  \bvol{20}~(5),  \pg{579--598}.

\bibitem[Benomar {\em et~al.\/}(2018)Benomar, Bazot, Nielsen, Gizon, Sekii,
  Takata, Hotta, Hanasoge, Sreenivasan \& Christensen-Dalsgaard]{Benomar2018}
{\sc \au{Benomar, O.}, \au{Bazot, M.}, \au{Nielsen, M.~B.}, \au{Gizon, L.},
  \au{Sekii, T.}, \au{Takata, M.}, \au{Hotta, H.}, \au{Hanasoge, S.},
  \au{Sreenivasan, K.~R.} \& \au{Christensen-Dalsgaard, J.}} \yr{2018}
  \at{Asteroseismic detection of latitudinal differential rotation in 13
  sun-like stars}.  \jt{Science}  \bvol{361}~(6408),  \pg{1231--1234}.

\bibitem[Bonnevier(1966)]{Bonnevier1966}
{\sc \au{Bonnevier, B.}} \yr{1966}  \at{Diffusion due to ion-ion collisions in
  a multicomponent plasma}.  \jt{Ark. Fys.}  \bvol{33},  \pg{255}.

\bibitem[Brillouin(1945)]{Brillouin1945}
{\sc \au{Brillouin, L.}} \yr{1945}  \at{A theorem of {L}armor and its
  importance for electrons in magnetic fields}.  \jt{Phys. Rev.}
  \bvol{67}~(7-8),  \pg{260--266}.

\bibitem[Chen {\em et~al.\/}(2017)Chen, Qin \& Liu]{Chen2017}
{\sc \au{Chen, Q.}, \au{Qin, H.} \& \au{Liu, J.}} \yr{2017}  \at{Photons,
  phonons, and plasmons with orbital angular momentum in plasmas}.  \jt{Sci.
  Rep.}  \bvol{7}~(1).

\bibitem[Davidson(2001)]{Davidson2001}
{\sc \au{Davidson, R.~C.}} \yr{2001} {\em Physics of Nonneutral Plasmas\/}.
  \publ{Imperial College Press}.

\bibitem[Dolgolenko \& Muromkin(2009)]{Dolgolenko2009}
{\sc \au{Dolgolenko, D.~A.} \& \au{Muromkin, Y.~A.}} \yr{2009}  \at{Plasma
  isotope separation based on ion cyclotron resonance}.  \jt{Physics-Uspekhi}
  \bvol{52}~(4),  \pg{345--357}.

\bibitem[Dolgolenko \& Muromkin(2017)]{Dolgolenko2017}
{\sc \au{Dolgolenko, D.~A.} \& \au{Muromkin, Y.~A.}} \yr{2017}  \at{Separation
  of mixtures of chemical elements in plasma}.  \jt{Phys. Usp.}
  \bvol{60}~(10),  \pg{994}.

\bibitem[Ellis {\em et~al.\/}(2005)Ellis, Case, Elton, Ghosh, Griem, Hassam,
  Lunsford, Messer \& Teodorescu]{Ellis2005}
{\sc \au{Ellis, R.~F.}, \au{Case, A.}, \au{Elton, R.}, \au{Ghosh, J.},
  \au{Griem, H.}, \au{Hassam, A.}, \au{Lunsford, R.}, \au{Messer, S.} \&
  \au{Teodorescu, C.}} \yr{2005}  \at{Steady supersonically rotating plasmas in
  the {M}aryland centrifugal experiment}.  \jt{Phys. Plasmas}  \bvol{12}~(5),
  \pg{055704}.

\bibitem[van Enk \& Nienhuis(1994)]{Enk1994}
{\sc \au{van Enk, S.~J.} \& \au{Nienhuis, G.}} \yr{1994}  \at{Spin and orbital
  angular momentum of photons}.  \jt{Europhys. Lett.}  \bvol{25}~(7),
  \pg{497--501}.

\bibitem[Faraday(1846)]{Faraday1846}
{\sc \au{Faraday, M.}} \yr{1846}  \at{On the magnetization of light and the
  illumination of magnetic lines of force}.  \jt{Phil. Trans. Roy. Soc. London}
   \bvol{136}~(1),  \pg{1--20}.

\bibitem[Fetterman \& Fisch(2008)]{Fetterman2008}
{\sc \au{Fetterman, A.~J.} \& \au{Fisch, N.~J.}} \yr{2008}  \at{Alpha
  channeling in a rotating plasma}.  \jt{Phys. Rev. Lett.}  \bvol{101}~(20),
  \pg{205003--}.

\bibitem[Fetterman \& Fisch(2010)]{Fetterman2010}
{\sc \au{Fetterman, A.~J.} \& \au{Fisch, N.~J.}} \yr{2010}  \at{Alpha
  channeling in rotating plasma with stationary waves}.  \jt{Physics of
  Plasmas}  \bvol{17}~(4),  \pg{042112}.

\bibitem[Franke-Arnold {\em et~al.\/}(2011)Franke-Arnold, Gibson, Boyd \&
  Padgett]{Franke-Arnold2011}
{\sc \au{Franke-Arnold, S.}, \au{Gibson, G.}, \au{Boyd, R.~W.} \& \au{Padgett,
  M.~J.}} \yr{2011}  \at{Rotary photon drag enhanced by a slow-light medium}.
  \jt{Science}  \bvol{333}~(6038),  \pg{65}.

\bibitem[Fresnel(1818)]{Fresnel1818}
{\sc \au{Fresnel, A.}} \yr{1818}  \at{Lettre d' {A}ugustin {F}resnel {\`a}
  {F}rançois {A}rago sur l' influence du mouvement terrestre dans quelques
  ph{\'e}nom{\`e}nes d'optique}.  \jt{Ann. Chim. Phys}  \bvol{9},  \pg{57--66}.

\bibitem[Gough(1986)]{Gough1986}
{\sc \au{Gough, W.}} \yr{1986}  \at{The angular momentum of radiation}.
  \jt{European J. Phys.}  \bvol{7}~(2),  \pg{81--87}.

\bibitem[Gueroult {\em et~al.\/}(2020)Gueroult, Rax \& Fisch]{Gueroult2020}
{\sc \au{Gueroult, R.}, \au{Rax, J.-M.} \& \au{Fisch, N.~J.}} \yr{2020}
  \at{Enhanced tuneable rotatory power in a rotating plasma}.  \jt{Phys. Rev.
  E}  \bvol{102}~(5),  \pg{051202(R)}.

\bibitem[Gueroult {\em et~al.\/}(2019{\natexlab{{\em a\/}}})Gueroult, Shi, Rax
  \& Fisch]{Gueroult2019a}
{\sc \au{Gueroult, R.}, \au{Shi, Y.}, \au{Rax, J.-M.} \& \au{Fisch, N.~J.}}
  \yr{2019{\natexlab{{\em a\/}}}}  \at{Determining the rotation direction in
  pulsars}.  \jt{Nat. Commun.}  \bvol{10}~(1),  \pg{3232}.

\bibitem[Gueroult {\em et~al.\/}(2019{\natexlab{{\em b\/}}})Gueroult, Zweben,
  Fisch \& Rax]{Gueroult2019}
{\sc \au{Gueroult, R.}, \au{Zweben, S.~J.}, \au{Fisch, N.~J.} \& \au{Rax,
  J.-M.}} \yr{2019{\natexlab{{\em b\/}}}}  \at{{E x B} configurations for
  high-throughput plasma mass separation: An outlook on possibilities and
  challenges}.  \jt{Phys. Plasmas}  \bvol{26}~(4),  \pg{043511}.

\bibitem[Götte {\em et~al.\/}(2007)Götte, Barnett \& Padgett]{Goette2007}
{\sc \au{Götte, J.~B.}, \au{Barnett, S.~M.} \& \au{Padgett, M.}} \yr{2007}
  \at{On the dragging of light by a rotating medium}.  \jt{Proc. R. Soc. A}
  \bvol{463}~(2085),  \pg{2185--2194}.

\bibitem[Jones(1976)]{Jones1976}
{\sc \au{Jones, R.~V.}} \yr{1976}  \at{Rotary aether drag}.  \jt{Proc. R. Soc.
  London, Ser. A}  \bvol{349}~(1659),  \pg{423--439}.

\bibitem[Klozenberg {\em et~al.\/}(1965)Klozenberg, McNamara \&
  Thonemann]{Klozenberg1965}
{\sc \au{Klozenberg, J.~P.}, \au{McNamara, B.} \& \au{Thonemann, P.~C.}}
  \yr{1965}  \at{The dispersion and attenuation of helicon waves in a uniform
  cylindrical plasma}.  \jt{J. Fluid Mech.}  \bvol{21}~(3),  \pg{545--563}.

\bibitem[Kolmes {\em et~al.\/}(2019)Kolmes, Ochs, Mlodik, Rax, Gueroult \&
  Fisch]{Kolmes2019}
{\sc \au{Kolmes, E.~J.}, \au{Ochs, I.~E.}, \au{Mlodik, M.~E.}, \au{Rax, J.-M.},
  \au{Gueroult, R.} \& \au{Fisch, N.~J.}} \yr{2019}  \at{Radial current and
  rotation profile tailoring in highly ionized linear plasma devices}.
  \jt{Phys. Plasmas}  \bvol{26}~(8),  \pg{082309}.

\bibitem[Kostyukov {\em et~al.\/}(2002)Kostyukov, Shvets, Fisch \&
  Rax]{Kostyukov2002}
{\sc \au{Kostyukov, I.~Y.}, \au{Shvets, G.}, \au{Fisch, N.~J.} \& \au{Rax,
  J.~M.}} \yr{2002}  \at{Magnetic-field generation and electron acceleration in
  relativistic laser channel}.  \jt{Physics of Plasmas}  \bvol{9}~(2),
  \pg{636--648}.

\bibitem[Krishnan {\em et~al.\/}(1981)Krishnan, Geva \&
  Hirshfield]{Krishnan1981}
{\sc \au{Krishnan, M.}, \au{Geva, M.} \& \au{Hirshfield, J.~L.}} \yr{1981}
  \at{Plasma centrifuge}.  \jt{Phys. Rev. Lett.}  \bvol{46}~(1),  \pg{36--38}.

\bibitem[Leach {\em et~al.\/}(2008)Leach, Wright, Götte, Girkin, Allen,
  Franke-Arnold, Barnett \& Padgett]{Leach2008}
{\sc \au{Leach, J.}, \au{Wright, A.~J.}, \au{Götte, J.~B.}, \au{Girkin,
  J.~M.}, \au{Allen, L.}, \au{Franke-Arnold, S.}, \au{Barnett, S.~M.} \&
  \au{Padgett, M.~J.}} \yr{2008}  \at{{\textquotedblleft}{A}ether
  {D}rag{\textquotedblright} and moving images}.  \jt{Phys. Rev. Lett.}
  \bvol{100}~(15),  \pg{153902}.

\bibitem[Lehner {\em et~al.\/}(1989)Lehner, Rax \& Zou]{Lehner1989}
{\sc \au{Lehner, T.}, \au{Rax, J.~M.} \& \au{Zou, X.~L.}} \yr{1989}  \at{Linear
  mode conversion by magnetic fluctuations in inhomogeneous magnetized
  plasmas}.  \jt{Europhys. Lett.}  \bvol{8}~(8),  \pg{759--764}.

\bibitem[Lehnert(1962)]{Lehnert1962}
{\sc \au{Lehnert, B.}} \yr{1962}  \at{Stabilization of flute disturbances by
  the {C}oriolis force}.  \jt{Phys. Fluids}  \bvol{5}~(6),  \pg{740}.

\bibitem[Lehnert(1971)]{Lehnert1971}
{\sc \au{Lehnert, B.}} \yr{1971}  \at{Rotating plasmas}.  \jt{Nucl. Fusion}
  \bvol{11}~(5),  \pg{485--}.

\bibitem[Liziakin {\em et~al.\/}(2021)Liziakin, Antonov, Usmanov, Melnikov,
  Timirkhanov, Vorona, Smirnov, Oiler, Kislenko, Gavrikov \&
  Smirnov]{Liziakin2021}
{\sc \au{Liziakin, G.}, \au{Antonov, N.}, \au{Usmanov, R.}, \au{Melnikov, A.},
  \au{Timirkhanov, R.}, \au{Vorona, N.}, \au{Smirnov, V.~S.}, \au{Oiler, A.},
  \au{Kislenko, S.}, \au{Gavrikov, A.} \& \au{Smirnov, V.~P.}} \yr{2021}
  \at{Experimental demonstration of plasma mass separation in a configuration
  with a potential well and crossed electric and magnetic fields}.  \jt{Plasma
  Phys. Control. Fusion}  \bvol{63}~(3),  \pg{032002}.

\bibitem[Mendon{\c{c}}a(2012{\natexlab{{\em a\/}}})]{Mendonca2012a}
{\sc \au{Mendon{\c{c}}a, J.~T.}} \yr{2012{\natexlab{{\em a\/}}}}  \at{Kinetic
  description of electron plasma waves with orbital angular momentum}.
  \jt{Phys. Plasmas}  \bvol{19}~(11),  \pg{112113}.

\bibitem[Mendon{\c{c}}a(2012{\natexlab{{\em b\/}}})]{Mendonca2012}
{\sc \au{Mendon{\c{c}}a, J.~T.}} \yr{2012{\natexlab{{\em b\/}}}}  \at{Twisted
  waves in a plasma}.  \jt{Plasma Phys. Control. Fusion}  \bvol{54}~(12),
  \pg{124031}.

\bibitem[Mendonca {\em et~al.\/}(2009)Mendonca, Ali \&
  Thid{\'{e}}]{Mendonca2009}
{\sc \au{Mendonca, J.~T.}, \au{Ali, S.} \& \au{Thid{\'{e}}, B.}} \yr{2009}
  \at{Plasmons with orbital angular momentum}.  \jt{Phys. Plasmas}
  \bvol{16}~(11),  \pg{112103}.

\bibitem[Ochs \& Fisch(2017)]{Ochs2017}
{\sc \au{Ochs, I.~E.} \& \au{Fisch, N.~J.}} \yr{2017}  \at{Particle orbits in a
  force-balanced, wave-driven, rotating torus}.  \jt{Phys. Plasmas}  \bvol{24},
   \pg{092513}.

\bibitem[Ohkawa \& Miller(2002)]{Ohkawa2002}
{\sc \au{Ohkawa, T.} \& \au{Miller, R.~L.}} \yr{2002}  \at{Band gap ion mass
  filter}.  \jt{Phys. Plasmas}  \bvol{9}~(12),  \pg{5116--5120}.

\bibitem[Padgett {\em et~al.\/}(2006)Padgett, Whyte, Girkin, Wright, Allen,
  \"{O}hberg \& Barnett]{Padgett2006}
{\sc \au{Padgett, M.}, \au{Whyte, G.}, \au{Girkin, J.}, \au{Wright, A.},
  \au{Allen, L.}, \au{\"{O}hberg, P.} \& \au{Barnett, S.~M.}} \yr{2006}
  \at{Polarization and image rotation induced by a rotating dielectric rod:
  {A}n optical angular momentum interpretation}.  \jt{Opt. Lett.}
  \bvol{31}~(14),  \pg{2205}.

\bibitem[Player(1976)]{Player1976}
{\sc \au{Player, M.~A.}} \yr{1976}  \at{On the dragging of the plane of
  polarization of light propagating in a rotating medium}.  \jt{Proc. R. Soc.
  A}  \bvol{349}~(1659),  \pg{441}.

\bibitem[Prasad \& Krishnan(1987)]{Prasad1987}
{\sc \au{Prasad, R.~R.} \& \au{Krishnan, M.}} \yr{1987}  \at{Rigid rotor
  equilibria of multifluid, neutral plasma columns in crossed electric and
  magnetic fields}.  \jt{Phys. Fluids}  \bvol{30}~(11),  \pg{3496}.

\bibitem[Rax {\em et~al.\/}(2015)Rax, Fruchtman, Gueroult \& Fisch]{Rax2015}
{\sc \au{Rax, J.~M.}, \au{Fruchtman, A.}, \au{Gueroult, R.} \& \au{Fisch,
  N.~J.}} \yr{2015}  \at{Breakdown of the {B}rillouin limit and classical
  fluxes in rotating collisional plasmas}.  \jt{Phys. Plasmas}  \bvol{22}~(9),
  \pg{092101}.

\bibitem[Rax \& Gueroult(2016)]{Rax2016}
{\sc \au{Rax, J.-M.} \& \au{Gueroult, R.}} \yr{2016}  \at{Rotation and
  instabilities for isotope and mass separation}.  \jt{J. Plasma Phys.}
  \bvol{82},  \pg{595820504}.

\bibitem[Rax {\em et~al.\/}(2017)Rax, Gueroult \& Fisch]{Rax2017}
{\sc \au{Rax, J.~M.}, \au{Gueroult, R.} \& \au{Fisch, N.~J.}} \yr{2017}
  \at{Efficiency of wave-driven rigid body rotation toroidal confinement}.
  \jt{Phys. Plasmas}  \bvol{24}~(3),  \pg{032504}.

\bibitem[Rax {\em et~al.\/}(2019)Rax, Kolmes, Ochs, Fisch \&
  Gueroult]{Rax2019a}
{\sc \au{Rax, J.~M.}, \au{Kolmes, E.~J.}, \au{Ochs, I.~E.}, \au{Fisch, N.~J.}
  \& \au{Gueroult, R.}} \yr{2019}  \at{Nonlinear ohmic dissipation in
  axisymmetric {DC} and {RF} driven rotating plasmas}.  \jt{Phys. Plasmas}
  \bvol{26}~(1),  \pg{012303}.

\bibitem[Rax \& Robiche(2010)]{Rax2010}
{\sc \au{Rax, J.-M.} \& \au{Robiche, J.}} \yr{2010}  \at{Theory of unfolded
  cyclotron accelerator}.  \jt{Physics of Plasmas}  \bvol{17}~(10),  \pg{--}.

\bibitem[Rax {\em et~al.\/}(2007)Rax, Robiche \& Fisch]{Rax2007}
{\sc \au{Rax, J.-M.}, \au{Robiche, J.} \& \au{Fisch, N.~J.}} \yr{2007}
  \at{Autoresonant ion cyclotron isotope separation}.  \jt{Phys. Plasmas}
  \bvol{14}~(4),  \pg{043102}.

\bibitem[Robiche \& Rax(2008)]{Robiche2008}
{\sc \au{Robiche, J.} \& \au{Rax, J.-M.}} \yr{2008}  \at{Relativistic
  wave-induced splitting of the {L}angmuir mode in a magnetized plasma}.
  \jt{Physical Review E}  \bvol{77}~(1),  \pg{016402}.

\bibitem[Ruseckas {\em et~al.\/}(2007)Ruseckas, Juzeli{\~u}nas, {\"O}hberg \&
  Barnett]{Ruseckas2007}
{\sc \au{Ruseckas, J.}, \au{Juzeli{\~u}nas, G.}, \au{{\"O}hberg, P.} \&
  \au{Barnett, S.~M.}} \yr{2007}  \at{Polarization rotation of slow light with
  orbital angular momentum in ultracold atomic gases}.  \jt{Phys. Rev. A}
  \bvol{76}~(5),  \pg{053822}.

\bibitem[Shinohara \& Horii(2007)]{Shinohara2007}
{\sc \au{Shinohara, S.} \& \au{Horii, S.}} \yr{2007}  \at{Initial trial of
  plasma mass separation by crossed electric and magnetic fields}.  \jt{Jap. J.
  App. Phys.}  \bvol{46}~(7R),  \pg{4276--}.

\bibitem[Shukla(2012)]{Shukla2012}
{\sc \au{Shukla, P.~K.}} \yr{2012}  \at{Twisted shear {A}lfv{\'{e}}n waves with
  orbital angular momentum}.  \jt{Phys. Lett. A}  \bvol{376}~(44),
  \pg{2792--2794}.

\bibitem[Shvets {\em et~al.\/}(2002)Shvets, Fisch \& Rax]{Shvets2002}
{\sc \au{Shvets, G.}, \au{Fisch, N.~J.} \& \au{Rax, J.-M.}} \yr{2002}
  \at{Magnetic field generation through angular momentum exchange between
  circularly polarized radiation and charged particles}.  \jt{Physical Review
  E}  \bvol{65}~(4),  \pg{046403--}.

\bibitem[Stenzel(2016)]{Stenzel2016a}
{\sc \au{Stenzel, R.~L.}} \yr{2016}  \at{Whistler waves with angular momentum
  in space and laboratory plasmas and their counterparts in free space}.
  \jt{Adv. Phys. X}  \bvol{1}~(4),  \pg{687--710}.

\bibitem[Stenzel(2019)]{Stenzel2019}
{\sc \au{Stenzel, R.~L.}} \yr{2019}  \at{Whistler modes excited by magnetic
  antennas: {A} review}.  \jt{Phys. Plasmas}  \bvol{26}~(8),  \pg{080501}.

\bibitem[Stenzel \& Urrutia(2014)]{Stenzel2014}
{\sc \au{Stenzel, R.~L.} \& \au{Urrutia, J.~M.}} \yr{2014}  \at{Magnetic
  antenna excitation of {W}histler modes. {II}. {A}ntenna arrays}.  \jt{Phys.
  Plasmas}  \bvol{21}~(12),  \pg{122108}.

\bibitem[Stenzel \& Urrutia(2015{\natexlab{{\em a\/}}})]{Stenzel2015a}
{\sc \au{Stenzel, R.~L.} \& \au{Urrutia, J.~M.}} \yr{2015{\natexlab{{\em
  a\/}}}}  \at{Helicon modes in uniform plasmas. {III}. {A}ngular momentum}.
  \jt{Phys. Plasmas}  \bvol{22}~(9),  \pg{092113}.

\bibitem[Stenzel \& Urrutia(2015{\natexlab{{\em b\/}}})]{Stenzel2015}
{\sc \au{Stenzel, R.~L.} \& \au{Urrutia, J.~M.}} \yr{2015{\natexlab{{\em
  b\/}}}}  \at{Helicon waves in uniform plasmas. {II}. {H}igh m numbers}.
  \jt{Phys. Plasmas}  \bvol{22}~(9),  \pg{092112}.

\bibitem[Stenzel \& Urrutia(2016)]{Stenzel2016}
{\sc \au{Stenzel, R.~L.} \& \au{Urrutia, J.~M.}} \yr{2016}
  \at{Trivelpiece-{G}ould modes in a uniform unbounded plasma}.  \jt{Phys.
  Plasmas}  \bvol{23}~(9),  \pg{092103}.

\bibitem[Stenzel \& Urrutia(2018)]{Stenzel2018}
{\sc \au{Stenzel, R.~L.} \& \au{Urrutia, J.~M.}} \yr{2018}  \at{Helicons in
  uniform fields. {II}. {P}oynting vector and angular momenta}.  \jt{Phys.
  Plasmas}  \bvol{25}~(3),  \pg{032112}.

\bibitem[Stix(1971)]{Stix1971}
{\sc \au{Stix, T.~H.}} \yr{1971}  \at{Some toroidal equilibria for plasma under
  magnetoelectric confinement}.  \jt{Phys. Fluids}  \bvol{14}~(3),  \pg{692}.

\bibitem[Stix(1992)]{Stix1992}
{\sc \au{Stix, T.~H.}} \yr{1992} {\em Waves in Plasmas\/}.  \publ{New York: AIP
  Press}.

\bibitem[Thaury {\em et~al.\/}(2013)Thaury, Guillaume, Corde, Lehe,
  Le~Bouteiller, Ta~Phuoc, Davoine, Rax, Rousse \& Malka]{Thaury2013}
{\sc \au{Thaury, C.}, \au{Guillaume, E.}, \au{Corde, S.}, \au{Lehe, R.},
  \au{Le~Bouteiller, M.}, \au{Ta~Phuoc, K.}, \au{Davoine, X.}, \au{Rax, J.~M.},
  \au{Rousse, A.} \& \au{Malka, V.}} \yr{2013}  \at{Angular-momentum evolution
  in laser-plasma accelerators}.  \jt{Physical Review Letters}
  \bvol{111}~(13),  \pg{135002--}.

\bibitem[Trivelpiece \& Gould(1959)]{Trivelpiece1959}
{\sc \au{Trivelpiece, A.~W.} \& \au{Gould, R.~W.}} \yr{1959}  \at{Space charge
  waves in cylindrical plasma columns}.  \jt{J. Appl. Phys.}  \bvol{30}~(11),
  \pg{1784--1793}.

\bibitem[Uberoi \& Das(1970)]{Uberoi1970}
{\sc \au{Uberoi, C.} \& \au{Das, G.~C.}} \yr{1970}  \at{Wave propagation in
  cold plasma in the presence of the {C}oriolis force}.  \jt{Plasma Phys.}
  \bvol{12}~(9),  \pg{661--684}.

\bibitem[Urrutia \& Stenzel(2015{\natexlab{{\em a\/}}})]{Urrutia2015a}
{\sc \au{Urrutia, J.~M.} \& \au{Stenzel, R.~L.}} \yr{2015{\natexlab{{\em
  a\/}}}}  \at{Helicon modes in uniform plasmas. {I}. {L}ow m modes}.
  \jt{Phys. Plasmas}  \bvol{22}~(9),  \pg{092111}.

\bibitem[Urrutia \& Stenzel(2015{\natexlab{{\em b\/}}})]{Urrutia2015}
{\sc \au{Urrutia, J.~M.} \& \au{Stenzel, R.~L.}} \yr{2015{\natexlab{{\em
  b\/}}}}  \at{Magnetic antenna excitation of whistler modes. {III}. {G}roup
  and phase velocities of wave packets}.  \jt{Phys. Plasmas}  \bvol{22}~(7),
  \pg{072109}.

\bibitem[Urrutia \& Stenzel(2016)]{Urrutia2016}
{\sc \au{Urrutia, J.~M.} \& \au{Stenzel, R.~L.}} \yr{2016}  \at{Helicon waves
  in uniform plasmas. {IV}. {B}essel beams, {G}endrin beams, and helicons}.
  \jt{Phys. Plasmas}  \bvol{23}~(5),  \pg{052112}.

\bibitem[Weng {\em et~al.\/}(2017)Weng, Zhao, Sheng, Yu, Luan, Chen, Yu,
  Murakami, Mori \& Zhang]{Weng2017}
{\sc \au{Weng, S.}, \au{Zhao, Q.}, \au{Sheng, Z.}, \au{Yu, W.}, \au{Luan, S.},
  \au{Chen, M.}, \au{Yu, L.}, \au{Murakami, M.}, \au{Mori, W.~B.} \& \au{Zhang,
  J.}} \yr{2017}  \at{Extreme case of {F}araday effect: magnetic splitting of
  ultrashort laser pulses in plasmas}.  \jt{Optica}  \bvol{4}~(9),  \pg{1086}.

\bibitem[Wilcox(1959)]{Wilcox1959}
{\sc \au{Wilcox, J.~M.}} \yr{1959}  \at{Review of high-temperature
  rotating-plasma experiments}.  \jt{Rev. Modern Phys.}  \bvol{31}~(4),
  \pg{1045--1051}.

\bibitem[Wisniewski-Barker {\em et~al.\/}(2014)Wisniewski-Barker, Gibson,
  Franke-Arnold, Boyd \& Padgett]{Wisniewski-Barker2014}
{\sc \au{Wisniewski-Barker, E.}, \au{Gibson, G.~M.}, \au{Franke-Arnold, S.},
  \au{Boyd, R.~W.} \& \au{Padgett, M.~J.}} \yr{2014}  \at{Mechanical {F}araday
  effect for orbital angular momentum-carrying beams}.  \jt{Opt. Express}
  \bvol{22}~(10),  \pg{11690--11697}.

\bibitem[Zweben {\em et~al.\/}(2018)Zweben, Gueroult \& Fisch]{Zweben2018}
{\sc \au{Zweben, S.~J.}, \au{Gueroult, R.} \& \au{Fisch, N.~J.}} \yr{2018}
  \at{Plasma mass separation}.  \jt{Phys. Plasmas}  \bvol{25}~(9),
  \pg{090901}.

\end{thebibliography}

\appendix

\section{Wave and particle spin and orbital angular momentum in a magnetised rotating plasma}
\label{Sec:Appendix_A}

In this appendix we briefly review how to separate SAM from OAM components for plasma waves and particles in light of the complementary dynamical and geometrical points of view. The up/down degree of freedom of the one half spin $\left( \hbar /2\right)$ of the electrons is not considered in this study and SAM refers to the cyclotron motion source of the cyclotron magnetic moment. 

The viewpoint adopted here is that separation of both wave's and classical particles' angular momentum into SAM and OAM components does not stem from a quantum point view, but instead is a simple generalisation of the first K\"{o}nig's theorem. K\"{o}nig's theorem states that the angular momentum of a system can be decomposed into an external orbital part and an internal part. This internal part is nowadays called spin part for waves and magnetised charges, even within a classical framework. The analysis of this separation for plasma waves and magnetised particles can be carried both from a geometrical point of view - through the analysis of the transformations properties of the wave field - and from a dynamical point of view - through the analysis of SAM and OAM waves and particles coupling. In this appendix we briefly review (\textit{i}) the canonical separation between SAM and OAM for a charged particle, then (\textit{ii}) the geometrical characterisation of SAM and OAM eigenvectors for a vector field and (\textit{iii}) previous results on the
resonant coupling of waves SAM with particles SAM and waves OAM with particles OAM.

\subsection{Canonical separation between SAM and OAM for a charged particle}

Consider a magnetised plasma column with a background magnetic field $\mathbf{B}$ directed along the $z$ axis and radially polarized by an electric field $\mathbf{E}$ sustaining a guiding center slow $E\times B$ rotation at angular velocity $\Omega $. A particle with charge $e$ and mass $m$ is described by (\textit{i}) its position $\mathbf{r}$ = $\mathbf{R}_{gc}$ + $\mathbf{\rho }_{L}$, where $\mathbf{R}_{gc}$ is the guiding center position and $\mathbf{\rho }_{L}$ the rotating Larmor radius, and (\textit{ii}) its velocity $\mathbf{v}$ = $\mathbf{V}_{d}$ + $\mathbf{v}_{c}$, where $V_{d}$ =
$\Omega R_{gc}$ is the guiding center $E\times B$ velocity and $v_{c}$ = $\omega _{c}\rho _{L}$ the cyclotron velocity. This configuration is illustrated on Fig.~\ref{Fig:Fig3}. The instantaneous angular momentum is given by $\mathbf{r}\times m\mathbf{v}$ and its average over the fast cyclotron motion is given by
\begin{equation}
\left\langle \mathbf{r}\times m\mathbf{v}\right\rangle =\left\langle \left[ 
\mathbf{R}_{gc}+\mathbf{\rho }_{L}\right] \times \left[ \mathbf{V}_{d}+\mathbf{v}_{c}\right] \right\rangle =m\Omega R_{gc}^{2}+m\omega _{c}\rho_{L}^{2}.
\end{equation}
Introducing the classical magnetic moment of the particle $\mu = m\omega _{c}^{2}\rho _{L}^{2}/2B$, the angular momentum $\left\langle\mathbf{r}\times m\mathbf{v}\right\rangle $ is then showed to be the sum of an orbital component $L_{z}$ plus a spin component $S_{z}$ associated with this magnetic moment,
\begin{eqnarray}
L_{z} &=&m\Omega R_{gc}^{2}, \\
S_{z} &=&2\frac{m}{e}\mu,  \label{samu}
\end{eqnarray}
where $e/(2m)$ is the classical gyromagnetic factor associated with the ratio of the magnetic moment to the angular momentum.

\begin{figure}
\begin{center}
\includegraphics[width = 6cm]{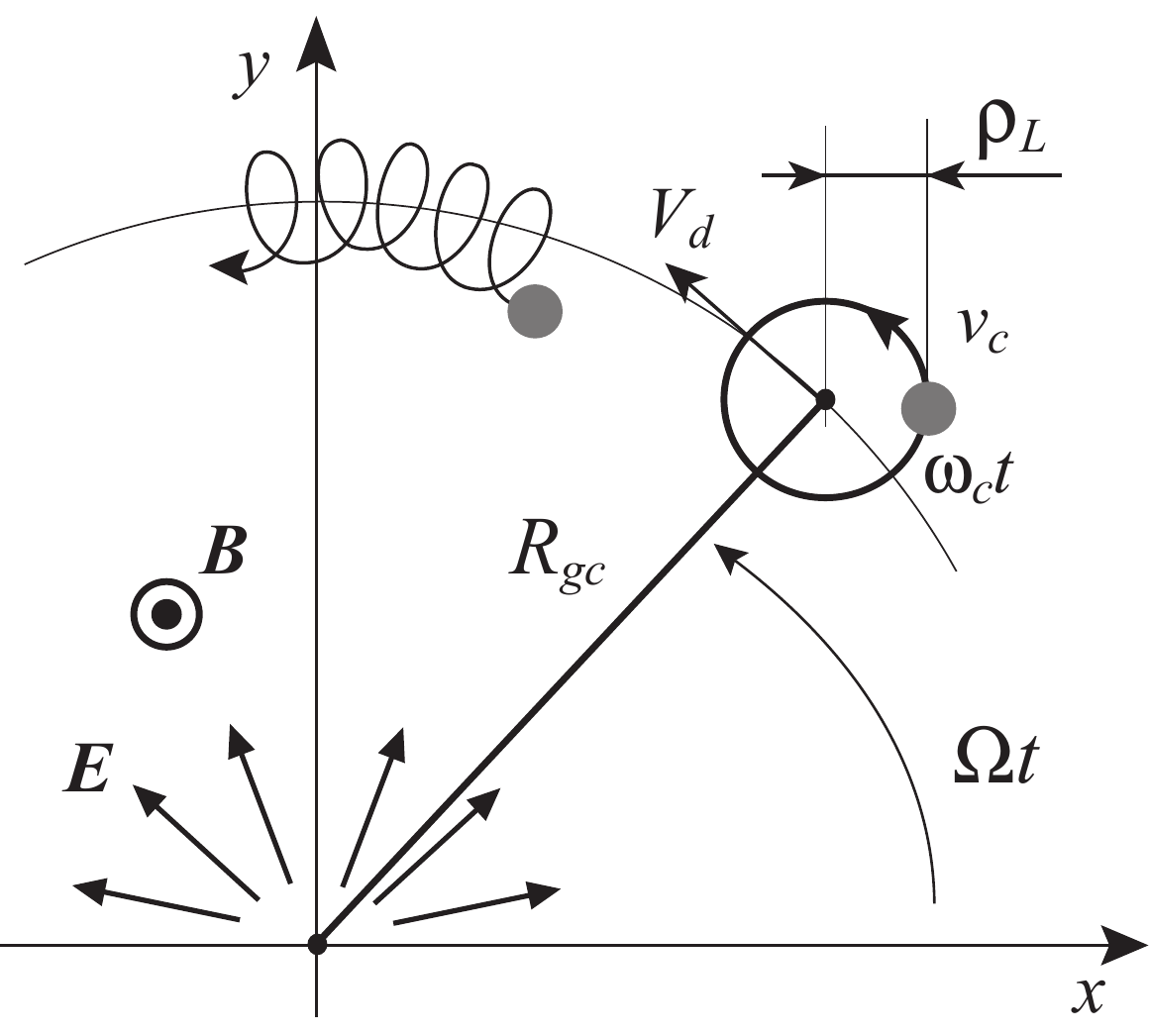}
\caption{Cyclotron (SAM) rotation and guiding center (OAM) $E\times B$ rotations in a magnetised rotating plasma column.}
 \label{Fig:Fig3}
\end{center}
\end{figure}

\subsection{Geometrical characterisation of SAM and OAM eigenvectors for a vector field}

Consider now a wave field $\mathbf{A}(\mathbf{r})\exp j\omega t$. From a purely geometrical point of view, the identification of (\textit{i}) linear momentum, (\textit{ii}) spin and (\textit{iii}) orbital momentum wave eigenstates can be guided by the analysis of the transformation properties of the wave under translations and rotations. 

Let us examine the change of this vector field $\mathbf{A}(\mathbf{r})$ under an active (change of the object $\mathbf{A}$) or a passive (change of the reference frame and coordinates used to describe $\mathbf{A}$) infinitesimal translation $\widehat{T}$ associated with the small vector $\delta \mathbf{r}$
\begin{equation}
\widehat{T}\mathbf{A}(\mathbf{r}) =\mathbf{A}\left( \mathbf{r}%
\pm \delta \mathbf{r}\right) =\mathbf{A}(\mathbf{r}) \pm \delta 
\mathbf{r.\nabla A}(\mathbf{r}),  \label{geo}
\end{equation}
with the minus and plus signs associated with the passive and active point of view, respectively. Eq.~(\ref{geo}) can be rewritten as a near identity transformation 
\begin{equation}
\widehat{T}\mathbf{A}(\mathbf{r}) =\left[ \mathbf{I}\pm j\delta 
\mathbf{r.}\widehat{\mathbf{P}}\right] \mathbf{A}(\mathbf{r}),
\end{equation}
where $\mathbf{I}$ \ is the identity operator and the linear momentum operator $\widehat{\mathbf{P}}$ is defined in the usual way as $\widehat{\mathbf{P}}=-j\bm{\nabla }$. The eigenvectors of this linear momentum operator $\widehat{\mathbf{P}}$ are the plane waves
\begin{equation}
\widehat{\mathbf{P}}\left( \exp j\bm{\beta }\cdot \mathbf{r}\right) =%
\bm{\beta }\left( \exp j\bm{\beta }\cdot \mathbf{r}\right).
\label{lme}
\end{equation}
These waves are also solutions of Maxwell-Amp\`{e}re and Maxwell-Faraday equations in an homogeneous linear dispersive plasma provided that $\bm{\beta }\left( \omega \right) $ verifies the dispersion relation. For a finite translation : $\mathbf{r}\pm \mathbf{a}$, the limit $N\rightarrow +
\infty $ of the iterated operator 
\begin{equation}
\left( \mathbf{I}\pm j\left( \mathbf{a}/N\right) \mathbf{\cdot }\widehat{\mathbf{P}}\right) ^{N}
\end{equation}
provides the expected Taylor expansion identity $\mathbf{A}\left( \mathbf{r}\pm \mathbf{a}\right) $ = $\left[ \exp \left( \pm \mathbf{a}\cdot \bm{\nabla }\right)
\right] \mathbf{A}(\mathbf{r}) $. 

Let us now examine the change of a vector field $\mathbf{A}(\mathbf{r}) $ under an active or passive  infinitesimal rotation $\widehat{R}$ associated with a small $\delta \theta $ rotation around an axis directed by $\mathbf{n}$ ($\mathbf{n}^{2}=1$) 
\begin{equation}
\widehat{R}\mathbf{A}(\mathbf{r}) =\pm \delta \theta \mathbf{n}\times \mathbf{A}\left( \mathbf{r}\pm \delta \theta \mathbf{n}\times \mathbf{r}\right) =\mathbf{A}(\mathbf{r}) \pm \delta \theta \left[ 
\mathbf{n}\times +\left( \mathbf{n}\times \mathbf{r}\right) \mathbf{.\nabla }\right] \mathbf{A}(\mathbf{r}).  \label{geo2}
\end{equation}
Eq.~(\ref{geo2}) can be rewritten as a near identity transformation displaying the separation between OAM and SAM operators 
\begin{equation}
\widehat{R}\mathbf{A}(\mathbf{r}) =\left[ \mathbf{I}\pm j\delta
\theta \mathbf{n\cdot }\left( \widehat{\mathbf{L}}+\widehat{\mathbf{S}}%
\right) \right] \mathbf{A}(\mathbf{r}).
\end{equation}
The OAM operator $\widehat{\mathbf{L}}$ \ and the SAM operator $\widehat{\mathbf{S}}$ are thus defined according to the usual relations 
\begin{eqnarray}
\widehat{\mathbf{L}} &=&\mathbf{r}\times \widehat{\mathbf{P}}=-j\left( 
\begin{array}{ccc}
y\frac{\partial }{\partial z}-z\frac{\partial }{\partial y} & 0 & 0 \\ 
0 & z\frac{\partial }{\partial x}-x\frac{\partial }{\partial z} & 0 \\ 
0 & 0 & x\frac{\partial }{\partial y}-y\frac{\partial }{\partial x}
\end{array}
\right), \\
\widehat{\mathbf{S}} &=&-j\mathbf{n}\times \mathbf{=}-j\left( 
\begin{array}{ccc}
0 & -n_{z} & n_{y} \\ 
n_{z} & 0 & -n_{x} \\ 
-n_{y} & n_{x} & 0
\end{array}
\right),
\end{eqnarray}
where these matrix representations are given on a cartesian basis $\left(\mathbf{e}_{x},\mathbf{e}_{y}\mathbf{,e}_{z}\right) $. Note that, since $\widehat{S}_{x}^{2}+\widehat{S}_{y}^{2}+\widehat{S}_{z}^{2}=2\mathbf{I}$ one recovers the angular momentum rule for a vector $\widehat{\mathbf{S}}^{2} = s\left( s+1\right) \mathbf{I}$ with $s = 1$.

The eigenvectors of this angular momentum operator $\widehat{\mathbf{L}}+\widehat{\mathbf{S}}$ are the vector spherical harmonics. However, since we are considering a magnetised plasma, vector spherical harmonics cannot be solution of Maxwell-Amp\`{e}re and Maxwell-Faraday equations. We have to restrict the transformations to translations along the magnetic field direction and rotations around the magnetic field direction, and the usefulness of the angular momentum operator is reduced to its $z$ component $\widehat{L}_{z}+\widehat{S}_{z}$. The eigenvectors of the OAM
operator $\widehat{L}_{z}$ are $\exp \pm jl\theta $ with $\theta $ the polar angle around the magnetic field, while the eigenvectors of the SAM operator $\widehat{S}_{z}$  are the L and R circularly polarised wave basis with eigenvalues $\pm 1$ and $\mathbf{e}_{z}$ with
the eigenvalue zero, that is
\begin{eqnarray}
\widehat{L}_{z}\left( \exp \pm jl\theta \right) &=&\pm l\left( \exp \pm
jl\theta \right),  \label{s4} \\
\widehat{S}_{z}\left( \frac{\mathbf{e}_{x}\pm j\mathbf{e}_{y}}{\sqrt{2}}\right) &=&\mp \left( \frac{\mathbf{e}_{x}\pm j\mathbf{e}_{y}}{\sqrt{2}}\right).  \label{s3}
\end{eqnarray}
Solutions of Maxwell-Amp\`{e}re and Maxwell-Faraday equations with a phase factor $\exp \pm jl\theta $ have a well defined OAM in a rotating magnetised plasma when the magnetic axis is also the rotation axis. Solutions of Maxwell-Amp\`{e}re and Maxwell-Faraday equations with a polarisation $\mathbf{e}_{x}\pm j\mathbf{e}_{y}$ have a well defined SAM. Solutions of Maxwell-Amp\`{e}re and Maxwell-Faraday equations with a phase factor $\exp \pm j\beta z$ have a well defined linear momentum.

\subsection{Resonant coupling}

The separation of OAM and SAM contents of waves outlined above can also be analysed through the coupling to the magnetised particles when resonant wave absorption takes place. To illustrate this point we go back to the magnetised plasma column depicted in Fig.~\ref{Fig:Fig3}. In this configuration the interplay between the wave's SAM and OAM components and particles' SAM and OAM can be summed up as follows. The wave's SAM (L or R polarisation) is coupled to the cyclotron SAM angular momentum ($2m\mu /e$) while the wave's OAM (helical phase fronts) is coupled to the OAM drift rotation of the guiding center around the axis of the plasma column. These separate coupling, which have previously been identified and studied within the frameworks of rotating thermonuclear trap~\citep{Rax2017}, dissipation in a fully ionized Brillouin flow~\citep{Rax2019a} and mass separation with rotating fields~\citep{Rax2016}, are recalled here.

Consider first a magnetised plasma column at rest and a transverse wave with $L_{z}=0$ and $S_{z}\neq 0$, that is to say with a simple oscillating phase factor $\exp j\left( \omega t-\beta z\right) $ and a polarisation component co-rotating with the cyclotron motion. The absorption of this wave by a population of charged particles with mass $m$ and charge $e$ leads to (\textit{i}) energy $\mathcal{E}$, (\textit{ii}) linear momentum $P_{z} = mv_{z}$ and (\textit{iii}) cyclotron angular momentum $S_{z} = mv_{c}^{2}/\omega _{c}$ exchange. Here $v_{c}$ is the cyclotron velocity around the magnetic field, $v_{z}$ the velocity along the magnetic field and $\omega _{c}$ the cyclotron frequency. For a collisionless resonant absorption through the classical Doppler shifted harmonic cyclotron resonance,
\begin{equation}
\omega -\beta v_{z}=n\omega _{c}, \qquad n\in \Bbb{Z},
\end{equation}
the absorption of an amount of energy $\delta \mathcal{E}$ by particles is associated both with a transfer $\delta P_{z} = m\delta v_{z}$ of linear momentum,
\begin{equation}
\delta P_{z}=\frac{\beta }{\omega }\delta \mathcal{E},  \label{h1}
\end{equation}
along the magnetic field lines and with a transfer $\delta S_{z}=mv_{c}\delta v_{c}/\omega _{c}$ of SAM defined by Eq.~(\ref{samu}), 
\begin{equation}
\delta S_{z}=2\frac{n}{\omega }\delta \mathcal{E},  \label{h4}
\end{equation}
around the magnetic field lines.

Consider now wave absorption in a magnetised rotating plasma column, and assume this time a wave with a helical phase factor $\exp j\left( \omega t-l\theta -\beta z\right)$ verifying the eigenvalue relations Eqs.~(\ref{s4}, \ref{lme}).  This wave has an OAM content such that $L_{z}/P_{z}=l/\beta\neq0$. In addition to the previous SAM coupling leading to $\delta S_{z}$, an OAM coupling takes place with the drift $E\times B$ rotation of the guiding center. The transfer of energy $\delta \mathcal{E}$ from this wave to a particle is associated with a transfer of OAM around the $z$ axis 
\begin{equation}
\delta L_{z}=\frac{l}{\omega }\delta \mathcal{E}.  \label{h3}
\end{equation}
Eqs.~(\ref{h1}, \ref{h4}, \ref{h3}) are simple consequences of energy,
linear momentum and angular momentum conservations
\begin{equation}
d\widehat{P}_{z}[\textit{waves} + \textit{particles}] \Big /dt= 0
\end{equation}
and
\begin{equation}
d\left( \widehat{L}_{z}+\widehat{S}_{z}\right)[\textit{waves} + \textit{particles}]\Big /dt= 0.
\end{equation}

Rather than examining these resonant exchange of linear and angular momentum, we have chosen in this study to consider the adiabatic interaction of waves and particles, that is to say the coupling between wave OAM and plasma $E\times B$ rotation on cold dispersion, as an analog to the coupling between wave SAM and plasma magnetisation on cold dispersion which yields the classical Faraday effect.

\end{document}